\documentclass{elsart}

\usepackage[american]{babel} 
\usepackage{cite}           
\usepackage{wrapfig}        
\usepackage[dvips]{epsfig}  
\usepackage{amsmath}        
\usepackage{amssymb}        
\usepackage{enumerate}      
\usepackage{xspace}         
\usepackage{rotating}

\journal{Nuclear Instruments \& Methods A}

\newcommand{\cuteval}{{\sc CutEval}\xspace}

\newcommand{\aman}{{\sf AMAN\-DA}\xspace}

\newcommand{\amax}{{\sf AMAN\-DA-B10}\xspace}

\newcommand{\amaii}{{{\sf AMAN\-DA}-II}\xspace}
\newcommand{\ice}{{\sf IceCube}\xspace}
\newcommand{\spase}{{\sf SPASE}\xspace}
\newcommand{\spasei}{{\sf SPASE-1}\xspace}
\newcommand{\spaseii}{{\sf SPASE-2}\xspace}

\newcommand{\cer}{Che\-ren\-kov\xspace}

\newcommand{\tld}{\hbox{\raise .15ex \hbox{$\sim$} \kern -.3em}}

\begin{document}

\begin{frontmatter}

\title{{Muon Track Reconstruction and Data Selection Techniques in \aman}
}


{\large The \aman Collaboration}\\
{\small
\begin{sloppypar}
\noindent
J.~Ahrens$^{11}$, 
X.~Bai$^{1}$, 
R.~Bay$^{9}$,
S.W.~Barwick$^{10}$, 
T.~Becka$^{11}$, 
J.K.~Becker$^{2}$,
K.-H.~Becker$^{2}$, 
E.~Bernardini$^{4}$, 
D.~Bertrand$^{3}$, 
A.~Biron$^{4}$
% ...
\end{sloppypar}

\vspace*{0.5cm} 

{\it 
\noindent
   (1) Bartol Research Institute, University of Delaware, Newark, DE 19716, USA
   \newline
   (2) Fachbereich 8 Physik, BU Wuppertal, D-42097 Wuppertal, Germany
   \newline
   (3) Universit\'e Libre de Bruxelles, Science Faculty, Brussels, Belgium
   \newline
   (4) DESY-Zeuthen, D-15738 Zeuthen, Germany
   \newline
   (5) Blackett Laboratory, Imperial College, London SW7 2BW, UK
   \newline
   (6) Dept.~of Technology, Kalmar University, S-39182 Kalmar, Sweden
   \newline
   (7) Lawrence Berkeley National Laboratory, Berkeley, CA 94720, USA
   \newline
   (8)~Dept.~of~Physics,~Pennsylvania~State~University,~University~Park,~PA~16802,~USA
   \newline
   (9) Dept. of Physics, University of California, Berkeley, CA 94720, USA
   \newline
   (10) Dept. of Physics and Astronomy, Univ. of California, Irvine, CA 92697, USA
   \newline
   (11) Institute of Physics, University of Mainz,  D-55099 Mainz, Germany
   \newline
   (12) Dept. of Physics, University of Maryland, College Park, MD 20742, USA
   \nopagebreak
   \newline
   (13) University of Mons-Hainaut, 7000 Mons, Belgium
   \nopagebreak
   \newline
   (14) Dept. de F\'{\i}sica, Universidad Sim\'on Bol\'{\i}var, Caracas, 1080, Venezuela
   \nopagebreak
   \newline
   (15) Dept. of Physics, University of Wisconsin, Madison, WI 53706, USA
   \nopagebreak
   \newline
   (16) Physics Dept., University of Wisconsin, River Falls, WI 54022, USA
   \nopagebreak
   \newline
   (17) Div. of High Energy Physics, Uppsala University, S-75121 Uppsala, Sweden   \nopagebreak
   \newline
   (18) Dept. of Physics, Stockholm University, SE-10691 Stockholm, Sweden
   \newline
   (19) Vrije Universiteit Brussel, Dienst ELEM, B-1050 Brussels, Belgium
   \newline
}

}

\begin{abstract}
The {\bf A}ntarctic {\bf M}uon {\bf A}nd {\bf N}eutrino {\bf D}etector {\bf
A}rray (AMANDA) is a high-energy neutrino telescope operating at the
geographic South Pole.  It is a lattice of photo-multiplier tubes buried
deep in the polar ice between $1500$\,m and $2000$\,m.  
The primary goal of this detector is to
discover astrophysical sources of high energy neutrinos. 
A high-energy muon neutrino coming through the earth
from the Northern Hemisphere can be identified by the secondary muon moving
upward through the detector.  

The muon tracks are reconstructed with a maximum likelihood method. It
models the arrival times and amplitudes of \cer photons registered
by the  photo-multipliers.
This paper describes the different methods of reconstruction,
which have been successfully implemented within \aman.
Strategies for optimizing the reconstruction performance
and rejecting background are presented.
For a typical analysis procedure the  direction
of tracks are reconstructed with about  $ 2^\circ $ accuracy.

\end{abstract}

\end{frontmatter}


\section{Introduction \label{sec:intro}}

\begin{figure}[hb!tp]
\vspace{.5cm}
\centerline{\epsfig{file=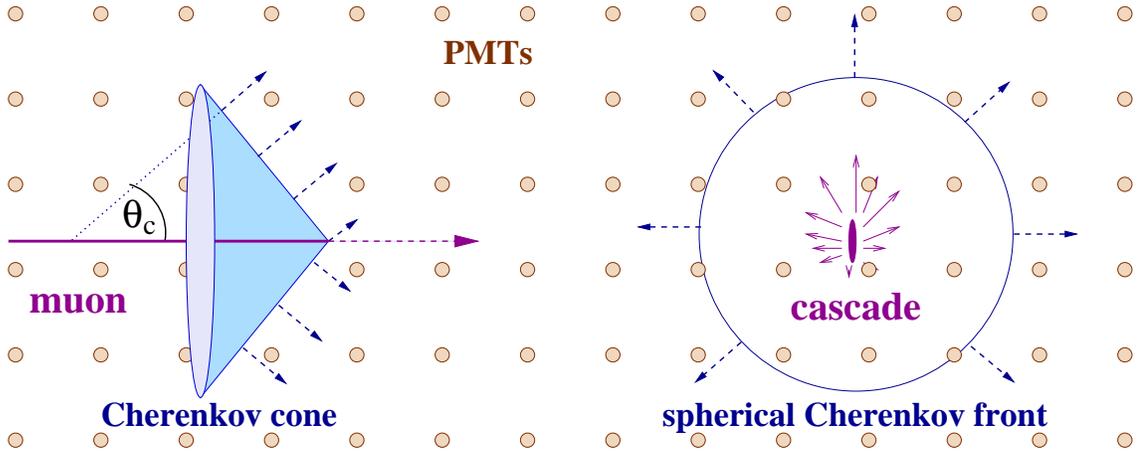,width=15cm}}
\vspace{.5cm}
\caption[Detection modes for high energy neutrinos]
{Detection modes of the \aman detector:
Left:  muon tracks induced by muon-neutrinos;
Right: Cascades from electron- or 
tau-neutrinos.
\label{fig:operationmodes}
}
\end{figure}

The  
 {\bf A}ntarctic {\bf M}uon {\bf A}nd {\bf N}eutrino {\bf D}etector
{\bf A}rray~\cite{ANDRES//:2001:A}, \aman, is a large volume  neutrino 
detector at the geographic South Pole.
It is a lattice of photo-multiplier tubes  (PMTs)
buried deep in the optically transparent polar ice.
The primary goal of this detector is to detect high-energy neutrinos
from astrophysical sources, and
determine their arrival time, direction and energy.
When a high-energy neutrino interacts in the polar ice
via a charged current reaction with a nucleon $N$:
\begin{equation}
\label{eq:ccreact}
\nu_\ell + N \rightarrow \ell + X~,
\end{equation}
it creates a hadronic cascade, $X$,
and a lepton, $\ell=e,\mu,\tau $.
These particles generate \cer photons,
which are detected by the  
PMTs.
Each lepton flavor generates a different
signal in the detector.
The two basic detection modes 
are sketched in figure~\ref{fig:operationmodes}.

A high-energy $\nu_\mu $ charged current interaction creates a  muon, which is 
nearly collinear with the neutrino direction; having
a mean deviation angle of
$ \psi = 0.7^\circ \,\times\, (E_{\nu}/\mathrm{TeV})^{-0.7}  $ \cite{LEARNED/MANNHEIM:2001:A},
which implies an accuracy requirement of $\lesssim 1^\circ$
for reconstructing the muon direction.

The high-energy muon emits a cone of \cer light
at a fixed angle $\theta_\mathrm{c}$. It is
determined by  $\cos \theta_\mathrm{c} = (n \cdot \beta)^{-1}$,
where $n \simeq 1.32$ is the index of
refraction in the ice.
For relativistic particles, $\beta \simeq 1 $, and 
$\theta_\mathrm{c} \approx 41^\circ $.
The direction of the muon is reconstructed from
the time and amplitude information of the PMTs
illuminated by the \cer cone.

Radiative energy loss processes generate secondary charged particles
along the muon trajectory, which also produce \cer radiation.  These
additional photons allow an estimate of the muon energy.  However, the
resolution is limited by fluctuations of these processes.  This
estimate is a lower bound on the neutrino energy, because it 
is based on the muon energy at 
the detector.  The interaction vertex
may be far outside the detector.

The $\nu_e$ and $\nu_\tau$ channels are different.
The electron from a $\nu_e $ will generate an 
electro-magnetic cascade, 
which is confined to a volume of a few cubic meters.
This cascade coincides with the hadronic cascade $X$
of the primary interaction vertex.
The optical signature is an expanding spherical shell of
\cer photons with a larger intensity in the 
forward direction.  The tau from a $\nu_\tau $
will decay immediately and also generate a 
cascade.
However, at energies $>1$\,PeV 
this  cascade and the vertex  are separated by several
tens of meters, connected by a single track. This signature
of two extremely bright  cascades is unique for
high-energy $\nu_\tau $, and it is called a 
{\em double bang event}~\cite{LEARNED/PAKVASA:1995:A}.

The measurement of cascade-like  events is restricted to interactions
close to the detector, thus requiring larger instrumented volumes
than for  $\nu_\mu $  detection.
Also the accuracy of the direction measurement is worse for 
cascades than for long muon tracks.
However, when the flux is diffuse, the
$\nu_e $ and $\nu_\tau $ channels also have clear advantages.
The backgrounds from atmospheric neutrinos are 
smaller.  The  energy resolution is significantly better
since the full energy is deposited in or near the detector.
The cascade channel is sensitive to all neutrino flavors
because the neutral current interactions also generate cascades.
In this paper we focus on the reconstruction of
muon tracks;  details on the reconstruction of cascades 
are described in \cite{AHRENS//:2002:D}.

The most abundant events in \aman are
atmospheric muons, created by cosmic rays 
interacting with the Earth's atmosphere.
At the depth of \aman their rate exceeds the rate of muons 
from atmospheric neutrinos by five orders of magnitude.
Since these muons are absorbed by the earth,
a muon track from the lower hemisphere
is a unique signature for a neutrino-induced
muon\footnote{Muon neutrinos  above $1\,$PeV 
are absorbed by the Earth. At these ultra-high-energies (UHE), however, the
the muon background from cosmic rays is small and UHE muons coming from 
the horizon and above  are most likely created by UHE neutrinos. The 
search for these UHE neutrinos is described in 
\cite{HUNDERTMARK//:2003:A,HUNDERTMARK//:2001:A}.}.
The reconstruction procedure must have good
angular resolution, good efficiency, and allow
excellent rejection of down-going atmospheric muons.

This paper describes the methods used to reconstruct
muon tracks recorded in the \aman experiment.
The  \amaii detector is introduced in section~\ref{sec:aman}.
The reconstruction algorithms and 
their implementation are described in sections~\ref{sec:algo}
to~\ref{sec:technical}.
Section~\ref{sec:reject} summarizes event classes for which the
reconstruction may fail and strategies to identify and eliminate such
 events.
The performance of the reconstruction procedure is shown in
section~\ref{sec:perform}.  We discuss possible 
improvements in
section~\ref{sec:ideas}.

\section{The \aman Detector\label{sec:aman}}

\begin{figure}[h!tb]
\centerline{\epsfig{file=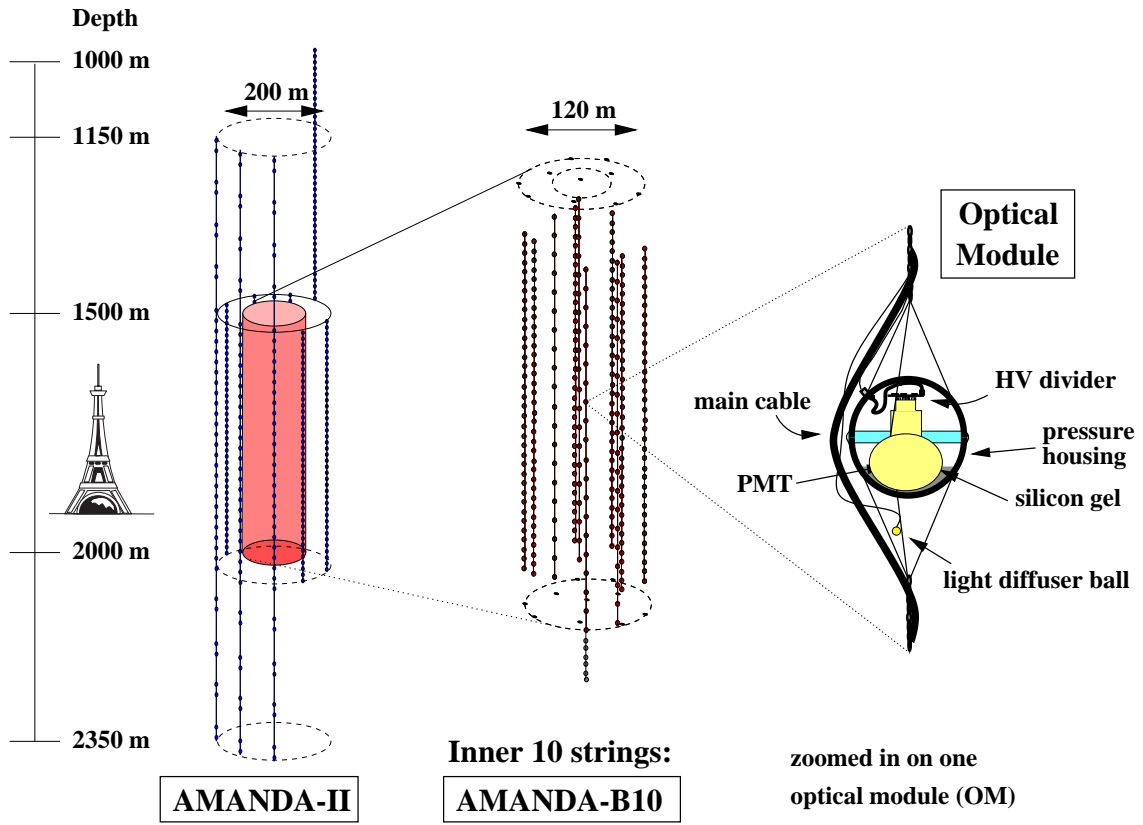,width=15cm}}
\caption[The  \aman detector]
{The  \amaii detector.
The scale is illustrated by the Eiffel tower at the left.
\label{fig:amashematic}
}
\end{figure}

The \amaii detector (see figure~\ref{fig:amashematic}) 
has been operating since January 2000 with
677 optical modules (OM) attached to 19 strings.  Most
of the OMs are located between 1500\,m and 2000\,m below the surface.
Each OM is a glass pressure vessel, which contains an 8-inch
hemispherical PMT and its electronics.  \amax\footnote{Occasionally in
the paper we will refer to this earlier detector instead of the full
\amaii detector.}, the inner core of 302 OMs on 10 strings, has been
operating since 1997.

One unique feature of
\aman is that it continuously measures atmospheric muons in coincidence with
the \emph{South Pole Air Shower Experiment} surface arrays \spasei and \spaseii \cite{DICKINSON//:2000:A}.  These muons are used to survey
the detector and calibrate the angular resolution
(see section~\ref{sec:perform} and
\cite{AHRENS//:2002:E,BAI//:2001:A}),
while providing \spase with
additional information for cosmic ray composition studies~\cite{RAWLINS:2001:A}.


The PMT signals are processed in a counting room at the surface of the
ice.  
The analog signals are amplified and sent to a majority logic trigger
 \cite{ANDRES//:2000:A}. There the pulses are discriminated and a 
trigger is formed if a minimum  number of hit PMTs are observed within a
time window of typically $2$\,$\mu $s. Typical trigger thresholds were $16$
hit PMT for \amax and $24$ for \amaii.
For each trigger the detector records the peak amplitude and
up to 16 leading and trailing edge times for each discriminated signal.
The time resolution achieved after calibration 
is ${\sigma}_t$ $\simeq 5$\,ns for 
the PMTs from the first 10 strings, which are read out via
coaxial or twisted pair cables.
For  the remaining PMTs, which are read out with optical fibers 
the  resolution is ${\sigma}_t$ $\simeq 3.5\,$ns.
In the cold environment of the deep ice the PMTs have low noise rates
of typically  $ 1\,$kHz.

The timing and amplitude calibration, the array geometry,
and the optical properties of the ice are determined
by illuminating the array with known optical pulses from
\emph{in situ} sources\cite{ANDRES//:2000:A}.
Time offsets are also determined from the
   response to through-going atmospheric muons \cite{COWEN//:2001:A}.

The optical absorption length in the ice is typically $ 110\,$m 
at 400\,nm with a strong wavelength dependence.
The effective scattering length at $400\, $nm is on average $\simeq 20\,$m.
It is defined as $\lambda_{\mathrm s}/(1-\langle\cos{\theta}_{\mathrm s}\rangle)$, where $\lambda_{\mathrm s}$ is
the scattering length and ${\theta}_{\mathrm s}$ is the scattering angle.
The ice parameters vary strongly with depth
due to horizontal ice layers, i.e., variations in the concentration
  of impurities which reflect 
past geological events and climate  changes
\cite{WOSCHNAGG//:1999:A,PRICE/WOSCHNAGG:2001:A,PRICE/WOSCHNAGG:2001:B,PRICE/WOS/CHI:2000:A,HE/PRICE/:1998:A,PRICE/BERGSTOEM:1997:A,ASKEBJER//:1997:A}.

\section{Reconstruction Algorithms\label{sec:algo}}

The muon track reconstruction algorithm is a maximum likelihood procedure.
Prior to reconstruction
simple pattern recognition algorithms,
discussed in section~\ref{sec:firstguess},
generate the initial estimates
required by the maximum likelihood reconstructions. 

\subsection{Likelihood Description}

The  reconstruction of an event
can be generalized to the problem of estimating a set of unknown parameters
$\{ {\mathbf{a}}\} $, e.g. track parameters, given a set of experimentally 
measured values $\{ {\mathbf{x}}\}$.
The parameters, $\{ {\mathbf{a}}\}$, are determined by maximizing the
likelihood 
${\mathcal L} (\mathbf{x} | \mathbf{a} )$ which for independent 
components $x_i$ of $\mathbf{x}$ reduces to
\begin{equation}
{\mathcal L} (\mathbf{x} | \mathbf{a} ) = \prod_i p(x_i | \mathbf{a})~,
\end{equation}
where $p(x_i | \mathbf{a})$ is the probability density function
(p.d.f.)
of observing the measured value $x_i$ for given values
of the parameters $\{ {\mathbf{a}}\} $ \cite{PDG:2000:A}.

\begin{figure}[ht]
\centerline{\epsfig{file=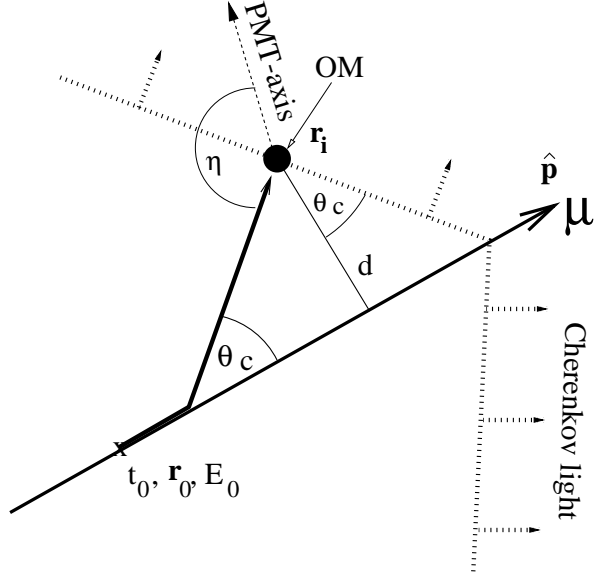,width=.49\textwidth}}
\caption[\cer geometry]
{\cer light front: definition of variables
\label{fig:cer_geo}
}
\end{figure}

To simplify the discussion we assume that the \cer radiation is generated
by a single infinitely long muon track (with $\beta = 1$) and forms a 
cone. It  is described by the following parameters:
\begin{equation}
\mathbf{a}  = (\mathbf{r_0} ,t_0, \mathbf{\hat{p}} , E_0)
\label{eq:onetrack}
\end{equation}
and illustrated in figure~\ref{fig:cer_geo}.  Here, $\mathbf{r_0}$ is
an arbitrary point on the track.  At time $t_0$, the muon passes
$\mathbf{r_0}$ with energy $E_0$ along a direction
$\mathbf{\hat{p}}$. The geometrical coordinates contain five degrees
of freedom.  Along this track, \cer photons are emitted at a fixed
angle $\theta_\mathrm{c}$ relative to $\mathbf{\hat{p}}$.  Within the
reconstruction algorithm it is possible to use a different coordinate system,
e.g. $\mathbf{a} =  (d, \eta, \dots )$.
The reconstruction is performed by minimizing
$ - \log (\mathcal{L}) $ with respect to $\mathbf{a}$.

The values $\{ {\mathbf{x}}\} $ presently recorded by  \aman 
are the time $t_i$ and duration  $TOT_i $ (\emph{Time Over Threshold}) 
of each PMT signal, as well as the peak amplitude $A_i$ of the largest
pulse in each PMT.  PMTs with no signal above threshold are also
accounted for in the likelihood function.  The hit times give the most
relevant information.  Therefore we will first concentrate on $p(t |
\mathbf{a}) $.

\subsubsection{Time Likelihood\label{sec:timelike}}

According to the geometry in figure~\ref{fig:cer_geo},
photons are 
expected to arrive at OM $i$ (at $\mathbf{r_i}$) at time
\begin{equation}
t_\mathrm{geo} = t_0 + \frac{\mathbf{\hat{p}} \cdot (\mathbf{r_i} - \mathbf{r_0})  
+ d \tan \theta_\mathrm{c}}{c_\mathrm{vac} }~,
\label{eq:defckovtime}
\end{equation}
with $c_\mathrm{vac}$ the vacuum speed of light\footnote{We note that 
equation~\ref{eq:defckovtime} neglects the effect that Cherenkov light
propagates with group velocity as pointed out in \cite{KUZMICHEV:2000:A}. 
It was shown in  \cite{PRICE/WOSCHNAGG:2001:A} that for AMANDA this 
approximation is justified.}.
It is convenient to define a relative arrival time, or \emph{time residual}
\begin{equation}
t_\mathrm{res} \equiv t_\mathrm{hit} - t_\mathrm{geo}~,
\label{eq:deftres}
\end{equation}
which is the difference between the observed hit time and the hit 
time expected for a ``direct photon'', a \cer photon that travels
undelayed directly from the muon to an OM without scattering.

\begin{figure}[htp]
\centerline{\epsfig{file=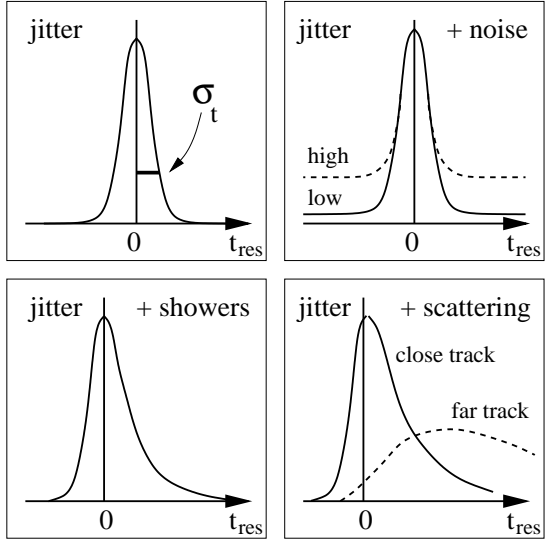,width=.45\textwidth}}
\caption[Arrival times]
{Schematic distributions of arrival 
times $t_\mathrm{tes}$ for different cases:  
Top left:  PMT jitter.
Top right: the effect of jitter and random noise.  
Bottom left: The effect of jitter and secondary cascades along the muon track.
Bottom right: The effect of jitter and scattering.
\label{fig:tres}
}
\end{figure}

In the ideal case, the distribution, $p(t_\mathrm{res} | \mathbf{a})$,
would be a delta function.  However, in a realistic experimental
situation this distribution is broadened and distorted by several
effects, which are illustrated in figure~\ref{fig:tres}.  The PMT
jitter limits the timing resolution ${\sigma}_t$.  Noise, e.g. dark
noise of the PMT, leads to additional hits which are random in time.  These
effects can generate negative $t_\mathrm{res}$ values, which 
would mimic unphysical causality violations.
Secondary radiative energy losses along the muon trajectory
create photons that arrive after the ideal \cer cone.
These processes are stochastic, and their relative photon
yield fluctuates.

In \aman, the dominant effect on photon arrival times is scattering in
the ice\footnote{In water detectors this effect is
neglected\cite{SPIERING//:1993:A} or treated as a small
correction\cite{CARMONA//:2001:A}.}.
The effect of scattering depends strongly on the distance,
$d$, of the OM from the track as illustrated in figure~\ref{fig:tres}.  
Since the PMTs have a non-uniform angular response, $p(t_\mathrm{res})$ 
also depends on the orientation, $\eta$, of the
OM relative to the muon track (see figure~\ref{fig:cer_geo}).
OMs facing away from the track can only
see light that scatters back towards the PMT face.  
On average this effect shifts $t_\mathrm{res} $ to  later times
and modifies the  probability of a hit.

The simplest time likelihood function is based on a  likelihood constructed
from $p_1$, the p.d.f.\ for arrival times of single photons $i$
at the locations of the hit OMs
\begin{equation}
 \mathcal{L}_\mathrm{time} = 
 \prod\limits_{i=1}^{N_\mathrm{hits}}  p_1(t_{\mathrm{res},i} | \mathbf{a} = 
   d_i, \eta_i , \dots ) ~.
\label{eq:timelike}
\end{equation}
Note that one OM may contribute to this product with several hits.
The function $ p_1(t_{\mathrm{res},i} | \mathbf{a} )$ is obtained from
the simulation of photon propagation through ice (see
section~\ref{sec:likeimplement}).  However, this description is
limited, because the electrical and optical signal channels can only
resolve multiple photons separated by a few $100$\,ns and $\simeq
10$\,ns, respectively.  Within this time window, only the arrival time
of the first pulse 
is recorded.

This first photon is usually less scattered than the average single
photon, which modifies the probability distribution of the detected hit time.
The arrival time distribution of the first
of $N$ photons  is given by 
\begin{equation}
\label{eq:defmultipe}
p_N^1(t_\mathrm{res})~=~ N \cdot  p_1 (t_\mathrm{res}) \cdot
\left ( \int_{t_\mathrm{res}}^\infty  p_1 (t) dt \right )^{(N-1)}
~=~ 
 N \cdot  p_1 (t_\mathrm{tres}) \cdot
\left ( 1 - P_1 (t_\mathrm{res}) \right )^{(N-1)}
\end{equation}
$P_1 $ is the cumulative distribution of the single photon p.d.f..
The function  $p_N^1(t_\mathrm{res})$ is called the multi-photo-electron 
({\em MPE}) p.d.f. and correspondingly defines $\mathcal{L}_\mathrm{MPE}$.

This concept can be extended to the more general case
of $p_N^k(t_\mathrm{res})$, the p.d.f.\  for the $k^\mathrm{th}$  
photon out of a total of $N$
to arrive at $t_\mathrm{res}$,
given by
\begin{equation} \label{eq:pnklike}
p_N^k(t_\mathrm{res}) = N \cdot 
\binom{N-1}{k-1}
  \cdot  p_1 (t_\mathrm{res}) \cdot
\left ( 1 - P_1 (t_\mathrm{res}) \right )^{(N-k)} \cdot
\left ( P_1 (t_\mathrm{res}) \right )^{(k-1)}
\end{equation}
$p_N^k(t_\mathrm{res})$
specifies the likelihood of arrival times of individual photoelectrons
for averaged time series of  $N$ photoelectrons.
With waveform recording the arrival times and amplitudes of
individual pulses can be resolved.

When the number of photoelectrons, $N$, is not measured precisely enough,
multi-photon information can be included via another method.
Instead of measuring $N$, 
the p.d.f.\ of the first photoelectron can be calculated
by convolving the MPE p.d.f.\ $p_N^1 (t,d) $ with the Poisson
probability $P_N^\mathrm{Poisson}(\mu)$, where $\mu$ is the mean 
expected number of photoelectrons as a function of the distance, $d$.
\begin{equation} 
p_\mu^1 (t_\mathrm{res})~=~ \frac{1}{N} 
\sum_{i=1}^\infty \frac{ \mu^i e^{-\mu }}{i!} \cdot p_i^1 (t_\mathrm{res}) 
~=~  
  \frac{\mu}{1 - e^{-\mu} } \cdot p_1(t_\mathrm{res}) 
        \cdot e^{ \displaystyle - \mu P_1(t_\mathrm{res})} 
\label{eq:psalike}
\end{equation}
This result is called the Poisson Saturated Amplitude (PSA) p.d.f.\
\cite{BOUCHTA:1998:A,PORRATA:1997:A}
 and correspondingly defines $\mathcal{L}_\mathrm{PSA} $.
The constant $N = 1 - e^{-\mu } $ renormalizes the p.d.f. to unity. 

The probability of  (uncorrelated) noise hits
is small.
They  are further suppressed by a \emph{hit cleaning}
procedure (section~\ref{sec:prepro}), which is applied before 
reconstruction. 
They 
are included in the likelihood function by 
simply adding a constant p.d.f.\ $p_0$.

\subsubsection{Hit and No-Hit  Likelihood \label{sec:lhit}}

The likelihood in the previous section relies only on the measured
arrival times of photons.  However, the topology of the hits is also
important.  PMTs with no hits near a hypothetical track or PMTs with hits far 
from the track are unlikely.

A likelihood   utilizing this  information
can be constructed as 
\begin{equation}
 \mathcal{L}_\mathrm{hit}  =  \prod_{i=1}^{N_\mathrm{ch}}
{P}^\mathrm{hit,i} \cdot \prod_{i=N_\mathrm{ch}+1}^{N_\mathrm{OM}}
{P}^\mathrm{no-hit,i} ,
\label{eq:defphitnohitlike}
\end{equation}
where $N_\mathrm{ch} $ is the number of hit OMs and $N_\mathrm{OM} $
the number of operational OMs.
The probabilities ${P}^\mathrm{hit}$ and ${P}^\mathrm{no-hit} $ 
of observing or not observing a hit depend
on the
track parameters $\mathbf{a}$. 
Additional hits due to
random noise are easily incorporated: 
$ P^\mathrm{no-hit} \to \tilde{P}^\mathrm{no-hit} 
\equiv  P^\mathrm{no-hit} \cdot  P^\mathrm{no-noise} $
and $  P^\mathrm{hit} \to \tilde{P}^\mathrm{hit} = 
1 - \tilde{P}^\mathrm{no-hit} $.

Assuming that the probability $P_1^\mathrm{hit}$ is known
for a single photon, the hit and  no-hit probabilities of OMs
for $n $ photons can be calculated:
\begin{equation}
\begin{array}{llcl}
&P^\mathrm{no-hit}_n & = & \left(1- P^\mathrm{hit}_1 \right)^n \\
\mbox{and} \quad &&&\\
&P^\mathrm{hit}_n & = & 1-P^\mathrm{no-hit}_n \\
&&=& 1-\left(1- P^\mathrm{hit}_1 \right)^n~.
\end{array}
\label{eq:defphitnohit}
\end{equation}
The number of photons, $n$, depends on $E_\mu$, the energy of the muon:
$n=n(E_\mu)$.  For a fixed track geometry,
the likelihood (equation~\ref{eq:defphitnohitlike}) can be used
to reconstruct the muon energy.

\subsubsection{Amplitude  Likelihood \label{sec:lamplitude}}

The peak amplitudes recorded by \aman can be fully
incorporated in the likelihood~\cite{MIOCINOVIC:2001:A},
which is particularly useful for energy reconstruction.
The likelihood can be written as 
\begin{equation}
 \mathcal{L} = \frac{W}{N_\mathrm{OM}} \cdot \prod_{i=1}^{N_\mathrm{OM}}
        w_i \cdot P_i(A_i ) ~,
\end{equation}
where  $ P_i(A_i ) $ is the probability that OM $i$
observes an amplitude $A_i $, with
$ A_i  = 0 $ for unhit OMs.
 $W$ and $w_i$ are weight factors,
which describe deviations of the individual OM and the 
total number of hit OMs from the expectation.
$P_i $ depends on the mean number
$  \mu $ of expected photoelectrons:
\begin{equation}
P_i (A_i) =  P^\mathrm{hit} \cdot 
 ( 1 - P_{i}^\mathrm{th}) \cdot
\frac{P(A_i,\mu )}{P(\langle A_i\rangle,\mu )}~.
\end{equation}
The probability $P_i (A_i) $ is normalized to the probability of 
observing the most likely amplitude $\langle A_i \rangle $.
$P_i^\mathrm{th} (\mu)  $ is the probability 
that a signal of
$\mu $ does not produce a pulse amplitude
above the discriminator threshold.
As before, $P^\mathrm{hit} = 1 - P^\mathrm{no-hit} $, where the 
no-hit probability is given by  Poisson statistics: 
$ P^\mathrm{no-hit} = \exp ( - \mu ) \cdot (1 - P^\mathrm{noise} ) $.
The probability of $A_i = 0 $ is a special case:
$ P_i(0) = P^\mathrm{no-hit} + P^\mathrm{hit} \cdot P_{i}^\mathrm{th}$.
Energy reconstructions based on this formulation of the likelihood
will be referred to as \emph{Full $E_{reco} $}.

An alternative energy reconstruction technique
(see section~\ref{sec:energyreco})
uses a neural net which is fed with energy sensitive parameters.

\subsubsection{Zenith Weighted (Bayesian) Likelihood\label{sec:weightedreco}}

Another extension of the 
likelihood~\cite{HILL:2001:A,Hill//:2002:A,COUSINS//:2002:A} incorporates 
external information
about the muon flux via Bayes' Theorem.
This theorem states that for two hypotheses $A$ and $B$, 
\begin{equation} 
P(A | B ) = \frac{P(B | A) P(A)}{P(B)}.
\label{eq:truestats}
\end{equation}
Identifying $A$ with the track parameters $\mathbf{a}$ and $B$ with the observations
$\mathbf{x}$, equation~\ref{eq:truestats} gives the probability that the inferred muon track
$\mathbf{a}$ was in fact responsible for the observed event $\mathbf{x}$.  
$P(\mathbf{x}|\mathbf{a})$ is 
the probability that $\mathbf{a}$, assumed to be true, would generate the event
$\mathbf{x}$ --- in other words, the likelihood described in
the previous sections. 
$P(\mathbf{a})$ is the prior probability
 of observing the
track $\mathbf{a}$; i.e., the relative frequencies of different muon tracks as
a function of their parameters. $P(\mathbf{x})$, which is independent 
of the track 
parameters $\mathbf{a}$, is a normalization constant which ensures that 
equation~\ref{eq:truestats} defines a proper probability.  Because the likelihood
is only defined up to an arbitrary constant factor, this normalization may
be ignored in the present context.

In order to obtain $P(\mathbf{a}|\mathbf{x})$, one thus has to determine
the prior probability distribution, $P(\mathbf{a})$, 
of how likely the various possible track directions are \emph{a priori}.
The reconstruction maximizes the product of the p.d.f.\ \emph{and} the prior.

The flux of muons deep underground is reasonably well known from previous 
experiments.  Any point source of muons
would be at most a small perturbation on the flux of penetrating
atmospheric muons and muons created by atmospheric neutrinos.
The most striking feature of the  background flux
from atmospheric muons  is the
strong dependence on zenith angle.
For vertically down-going tracks it exceeds the flux from neutrino 
induced muons by about 5 orders of magnitude but becomes negligible for
up-going tracks.
This  dependence, which is modeled 
by  a Monte Carlo calculation\cite{AHRENS//:2002:C},
acts as a zenith dependent weight to the different muon 
hypotheses, $\mathbf{a}$.
With this particular choice, some
tracks, which would otherwise reconstruct as up-going,
reconstruct as down-going tracks.  This greatly
reduces the rate at which penetrating atmospheric muons are 
mis-reconstructed as up-going 
neutrino events \cite{DEYOUNG:2001:A}.
In principle, a more accurate prior could be used.
It would need  to include 
the depth and energy dependence of the
atmospheric muons as well as the angular dependence of 
atmospheric neutrino induced muons.



Upon completion of this work, we learned that this technique was
developed independently by the NEVOD neutrino detector
collaboration\cite{AYNUTDINOV//:2001:A} who were able to extract an 
atmospheric neutrino
from a background of $10^{10}$ atmospheric muons in a small ($6 \times 6
\times 7.5 \mathrm{m}^3$)  surface detector.

\subsubsection{Combined Likelihoods\label{sec:combinedreco}}

The likelihood function $\mathcal{L}_\mathrm{time}$ of the
hit times is the most important for track reconstruction.
However, it is useful to include other information
like the hit probabilities.
The combined p.d.f.\ from Equations~\ref{eq:defmultipe} 
and~\ref{eq:defphitnohitlike} is
\begin{equation}
\mathcal{L}_\mathrm{MPE\oplus P^\mathrm{hit}P^\mathrm{no-hit}}=
\mathcal{L}_\mathrm{MPE}~ \cdot~ \left ( \mathcal{L}_\mathrm{hit} 
\right )^w~,
\label{eq:defcombined}
\end{equation}
which is particularly effective.  Here $w$ is an optional weight factor
 which allows the adjustment of the relative
weight of the two likelihoods.
This likelihood is sensitive not only to the track geometry 
but also to the energy of the muon.

As discussed in section~\ref{sec:weightedreco}, the  zenith angle dependent 
prior function, $P(\theta)$,
can  be included as a multiplicative factor. This  combination
\begin{equation}
        \mathcal{L}_\mathrm{Bayes} = P(\theta ) \cdot  
\mathcal{L}_\mathrm{time}~
\end{equation}
has  been used in the analysis 
of atmospheric neutrinos~\cite{AHRENS//:2002:C}.
However, all of these improved likelihoods
are limited by the underlying model assumption 
of a single  muon track.

\subsection{Likelihood Implementation \label{sec:likeimplement}}

The actual implementation of the likelihoods requires detailed
knowledge of the photon propagation in the ice.
On the other hand, efficiency considerations
and numeric problems favor a simple and robust method.

The photon hit
probabilities and arrival time distributions are simulated as functions
of all relevant parameters with a dedicated Monte Carlo simulation and
archived in large look-up tables.
This simulation  is described 
in~\cite{AHRENS//:2002:C,MIOCINOVIC:2001:A,KARLE:1999:C}.

The \aman Collaboration has followed different strategies for incorporating
this data into the reconstruction. 
In principle the probability density functions
are taken directly from these archives. 
However, one has to face several technical
difficulties due to the memory requirements of the archived tables, as
well as numeric problems related to the normalization of interpolated
bins and the calculation of multi-photon likelihoods.

Alternatively, one can simplify the model and 
parametrize these archives with analytical functions,
which depend only on a reduced set of parameters.
Comparisons of two independent
parametrizations~\cite{BOUCHTA:1998:A,WIEBUSCH:1999:A}
show that the direct and parametrized approaches yield
similar results in terms of efficiency.
This indicates that the parametrization
itself is not limiting the reconstruction
quality; rather, as mentioned earlier, the reconstruction is limited by
the assumptions of the model being fit.  Therefore, we will
concentrate on only one parametrization.

\subsubsection{Analytical Parametrization\label{sec:trackreco}\label{sec:pandelpar}}

A simple parametrization of the arrival time distributions
can be achieved with the following function, which we 
call \emph{Pandel function}.
It is a gamma distribution and its usage is motivated by
an analysis of laser light signals in the BAIKAL experiment
\cite{PANDEL:1996:A}.
There, it was found that for the case of an
isotropic, monochromatic and point-like light source,
$p_1(t_\mathrm{res})$ can be expressed in the form 
\begin{eqnarray}
p(t_\mathrm{res}) &~\equiv~ & \frac{1}{N(d)}\frac{\tau^{-(d/\lambda)}\cdot 
t_\mathrm{res}^{(d/\lambda-1)}}{\Gamma(d/\lambda)}\cdot 
e^{\displaystyle - \left(  t_\mathrm{res} \cdot \left ( \frac{1}{\tau} + 
\frac{c_\mathrm{medium}}{\lambda_a}
\right ) + \frac{d}{\lambda_a }
\right)} ~, \label{eq:defpandel} \\
N(d) &~=~& e^{-d/\lambda_a}\cdot 
\left(1+\frac{\tau \cdot c_\mathrm{medium}}{\lambda_a} \right)^{-d/\lambda}~,
\label{eq:normpandel}
\end{eqnarray}
without special assumptions on the actual optical parameters.
Here, $c_\mathrm{medium} = c_\mathrm{vac}/ n $ is the speed of light in ice, 
$\lambda_a $ the 
absorption length, $\Gamma(d/\lambda) $ the Gamma function
and $N(d)$ a normalization factor, which is given by equation~\ref{eq:normpandel}.
This formulation
 has free parameters $\lambda$ and $\tau$,
which are unspecified functions of the distance $d$ and the other
geometrical parameters.   They 
are empirically determined by a Monte Carlo 
model.

The Pandel function has some convenient mathematical properties: 
it is normalized, it is easy to compute,
and it can be integrated analytically 
over the time, $t_\mathrm{res}$, which simplifies the 
construction of the multi-photon (MPE) time p.d.f..
For small distances the function
has a pole at $t=0$ corresponding to a high 
probability of an unscattered photon. Going to larger values of $d$,
longer delay times become more likely. For distances larger than 
the critical value $d = \lambda$, the 
power index to $t_\mathrm{res} $ changes sign,
reflecting that 
the probability of  undelayed photons vanishes: essentially
all photons are delayed due to scattering.

The large freedom in the choice of the two parameter functions
$\tau$ (units of time) and $\lambda$ (units of length) and the overall
reasonable behavior is the motivation to use this function to parametrize
not only the time p.d.f.\ for point-like sources, but also
for muon tracks \cite{WIEBUSCH:1999:A}.
The Pandel function is fit to the distributions of delay times
for fixed distances $d$ and angles  $\eta$ (between the 
PMT axis and the \cer cone). These distributions are previously 
obtained from a detailed photon propagation 
Monte Carlo for the \cer light from muons.
The free fit parameters are $\tau $, $\lambda$, $\lambda_a$ and 
the effective distance $d_\mathrm{eff}$, which will be introduced next.

\begin{figure}[htpb]
\centerline{
\epsfig{file=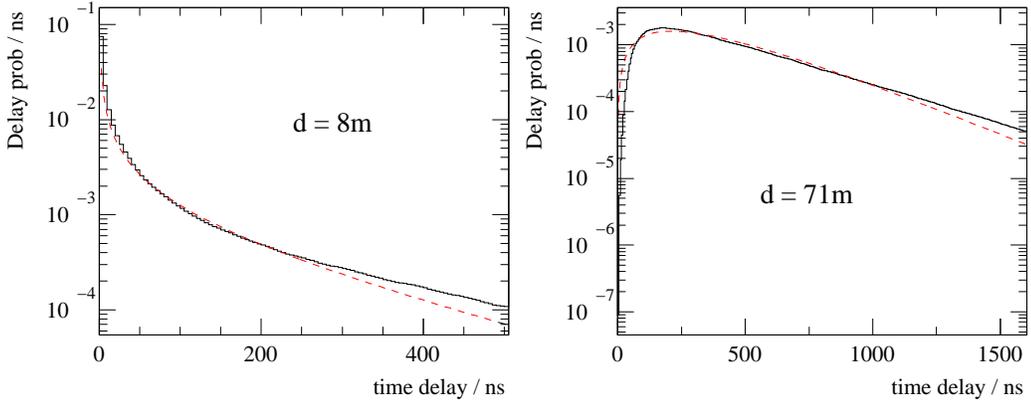,width=0.85\textwidth}
}
\caption{Comparison of the parametrized Pandel function (dashed curves) with the 
detailed simulation (black histograms) at two distances $d$ from the muon track.
\label{fig:pandelcompare}}
\end{figure}

When investigating the fit results as a function of $d$ and angle $\eta$ 
(see figure~\ref{fig:cer_geo}),
we observe that already for a simple ansatz 
of  constant $\tau$, $\lambda$ and $\lambda_a$
the optical properties in \aman are described sufficiently  well within
typical distances.
The dependence on $\eta$ is described by an effective distance
$d_\mathrm{eff}$ which replaces $d$ in equation~\ref{eq:defpandel}.
This means that the time delay distributions for backward illumination
of the PMT is found to be similar to a head-on illumination
at a larger distance.
The following parameters are obtained for a specific ice model, and
are currently used in the reconstruction:
\begin{equation}
\renewcommand{\arraystretch}{0.9}
\begin{array}{lcl}
\tau &=& 557\,\mbox{ns} \\
\lambda &=& 33.3\,\mbox{m}\\
\lambda_a &=& 98\,\mbox{m}  \\
\end{array}
\qquad
\begin{array}{lclcl}
 d_\mathrm{eff} &=& a_0 + a_1 \cdot d \\
  a_1 &=&  0.84    \\
  a_0 &=& 3.1\,\mbox{m} - 3.9\,\mbox{m} \cdot \cos( \eta) 
+ 4.6\,\mbox{m} \cdot \cos^2 (\eta)
\\
\end{array}
\renewcommand{\arraystretch}{1.0}
\label{eq:pandelpar}
\end{equation}
A comparison of the results from this parametrization 
with the full simulation is shown in figure~\ref{fig:pandelcompare} for 
two extreme distances.
The simple approximation describes the behavior
of the full simulation reasonably well.  
However, this simple overall description 
has a limited accuracy, especially for $ d\approx \lambda$ (not shown).
Reconstructions, based on the Pandel function with different ice models, 
and a generic reconstruction, that uses the full simulation results, 
yield similar results.
These comparisons  indicate that the results of the
reconstruction do not critically depend on the fine tuning of the underlying
ice models, and it justifies the use of the above simple model.

\subsubsection{Extension to realistic PMT signals \label{sec:upandel}}

Although the Pandel function is the basis of a simple normalized
likelihood, it has several deficiencies.  It is not defined for
negative $t_\mathrm{res}$, it ignores PMT jitter, and it has a pole at
$t_\mathrm{res} =0$, which causes numerical difficulties.  These
problems can be resolved by convolving the Pandel function,
equation~\ref{eq:defpandel}, with a Gaussian, which accounts for the PMT
jitter.  Unfortunately such convolution 
requires significant computing time.

Instead the Pandel function is modified by extending it to negative
times, $t_\mathrm{res} < 0 $, with a (half) Gaussian of width 
$\sigma_{g}$.
The effects of PMT jitter are only relevant
for small values of $t_\mathrm{res}$.
For times  $t_\mathrm{res} \ge t_1$ the original function is used,
 and the two parts are connected by a spline
interpolation (3rd order polynomial). 
The result, 
$\hat P (t_\mathrm{res})$, is 
called \emph{upandel function}.

Using $ t_1 = \sqrt{2\pi} \cdot \sigma_{g}$ and
requiring further a smooth interpolation and the normalization to be 
unchanged,
the polynomial coefficients $a_j $ and  normalization of the Gaussian
$ N_{g} $ can be calculated
analytically  \cite{WIEBUSCH:1999:A}.
The free parameter $\sigma_{g}$  includes
all timing  uncertainties, not just the PMT jitter. 
Good reconstruction results are achieved
for a large range  $10\,\mbox{ns} \le \sigma_{g} \le 20\,$ns.

\subsubsection{$P^\mathrm{hit}P^\mathrm{no-hit}$ 
        Parametrization \label{sec:phitreco}}

The normalization $N(d)$ in equation~\ref{eq:normpandel}
is used to construct a hit probability function, $P_\mathrm{hit}$.
The function $P_n^\mathrm{hit} $ with $ P_1^\mathrm{hit} \equiv N(d) $, 
is fit to the hit 
probability determined by the full \aman detector simulation,
as a function of distance,  orientation and muon energy. 
The free parameters are
the Pandel parameters $\tau $ and $\lambda $, $\lambda_a $,
 $\hat d$  and $\tilde n$. The effective distance  $\hat d$, is 
similar to the effective distance in the Pandel parametrization.
We define $\tilde n$  as the power index 
of  equation~\ref{eq:defphitnohit},  which
corresponds to an effective number of photons.
It is important to understand
that $N(d)$ is not a hit probability and $\tilde n$ is not just a 
number of photons.
They are  constructs, that are calibrated with a Monte Carlo 
simulation.
Technically, the power index, $\tilde n \equiv N $ in 
equation~\ref{eq:defphitnohit}, 
factorizes into  $\tilde n  (\eta, E_\mu )  = 
\epsilon (\eta,E_\mu) \cdot n(E_\mu )$.
The variable $n$, where $n=n(E)$, is related 
to the
number of photons  incident on the PMT and its absolute
efficiency.
The factor  $ \epsilon (\eta,E_\mu) $ 
is related to the orientation dependent  PMT  sensitivity 
but also accounts for the
 energy dependent angular emission profile of photons with respect 
to the bare muon.

\subsubsection{Energy Reconstruction\label{sec:energyreco}}

The reconstruction of the track geometry is a search for five
parameters.  If the muon energy is added as a fit parameter, the
minimization is significantly slower.  Therefore, the energy
reconstruction is performed in two steps.  First, the track geometry
is reconstructed without the energy parameter.  Then these geometric
parameters are used in an energy reconstruction, that only determines
the energy.
However, if the time likelihood utilizes amplitude information,
e.g.~ in the combined likelihood, equation~\ref{eq:defmultipe},
or the PSA likelihood, equation~\ref{eq:psalike},
the track parameters  also depend on the energy. 
In this case the energy and geometric parameters must
be reconstructed together.
Currently, three different approaches are used to reconstruct the muon energy.
They are compared in section~\ref{sec:energyperform}.

\begin{enumerate}
\item
The simplest method utilizes the 
$P^\mathrm{hit}P^\mathrm{no-hit}$ reconstruction
(see sections \ref{sec:lhit} and \ref{sec:phitreco}).

\item
The \emph{Full $E_{reco} $} method (see section~\ref{sec:lamplitude})
models the measured amplitudes in a likelihood for reconstructing the
energy.  This algorithm performs better, but it is more dependent on
the quality of the amplitude calibration of the OMs.

\item
An alternative way to measure the energy is based on a neural network
\cite{GEENEN//:2002A}.
The neural network uses 6-6-3-1 and 6-3-5-1 feed-forward architecture 
for \amax  and
\amaii, respectively.  The energy 
correlated variables which are used as input are
the mean of the measured amplitudes (ADC), 
the mean and RMS of the arrival times (LE) or  pulse durations (TOT), 
the total number of signals,
the number of OMs hit
and the number of OMs with exactly one hit.

\end{enumerate}

Less challenging than a full reconstruction,
a lower energy threshold is determined by requiring
a  minimum number of hit OMs. The number of hit OMs  is
correlated with the energy of the muon.
Since celestial neutrinos are believed to have a substantially 
harder spectrum than atmospheric neutrinos, an excess of high multiplicity
events would indicate that a hard celestial source exists.
Values for this parameter determined from \aman data already set
a tight upper limit on the diffuse flux of high-energy celestial neutrinos
\cite{AHRENS//:2003:A}.

\subsubsection{Cascade Reconstruction \label{sec:otherrecos}}

The reconstruction of \emph{cascade like events} 
is described in detail elsewhere 
\cite{AHRENS//:2002:D}.
The basic approach is similar to the track reconstruction.
It assumes events form a point light source with photons 
propagating spherically outside 
with a higher intensity  in the forward direction.
The cascade reconstruction also uses the Pandel function (see 
equation~\ref{eq:defpandel}) with parameters that are specific for cascades.

In several muon analyses, a cascade fit is used as a competing model. 
In cases where the cascade fit
achieves a better likelihood than the track reconstruction, 
the track hypothesis is rejected.
In particular this is used as a selection criterion 
to reject background events which are mis-reconstructed due to 
bright secondary cascades.


\section{First Guess Pattern Recognition \label{sec:firstguess}}

The likelihood reconstructions need an initial track hypothesis to
start the minimization.  The initial track is derived from \emph{first
guess methods}, which are fast analytic algorithms that do not require
an initial track.

\subsection{Direct Walk \label{sec:dirwalk}}

A very efficient \emph{first guess method} is the \emph{direct walk}
algorithm.  It is a pattern recognition algorithm based on carefully
selected hits, which were most likely caused by direct photons.

The four step procedure starts by selecting \emph{track elements}, the
straight line between any two hit OMs at distance $d$, which are hit
with a time difference
\begin{equation}
|\Delta t| < \frac{d}{c_{vac}}+30\,\mbox{ns} \mbox{~~with~~} d>50\,\mbox{m}~.
\label{eq:dwone}
\end{equation}
The known positions of the OMs define the track element direction
$(\theta, \phi)$.  The vertex position $(x, y, z)$ is taken at the
center between the two OMs.  The time at the vertex $t_0$ is defined
as the average of the two hit times.

In a next step, the number of \emph{associated hits} (AH) are
calculated for each track element.  Associated hits are those with
$-30\,\mbox{ns} < t_\mathrm{res} < 300\,$ns and
$d < 25\,\mbox{m} \cdot 
 (t_\mathrm{res} + 30)^{1/4}$
($t$ in ns), 
where $d$ is the distance between hit OM and track element and
$t_\mathrm{res}$ is the \emph{time residual}, which is defined in 
Equation~\ref{eq:deftres}.
After selecting these associated hits, track elements of poor
quality are rejected by requiring:
$
N_\mathrm{AH} \ge 10 \mbox{~and~}
\sigma_L \equiv \sqrt{(\frac{\displaystyle 1}{\displaystyle N_\mathrm{AH}}
\sum_i(L_i-\langle L\rangle)^2)} \ge 20\,\mbox{m}
$.
Here, the ``lever arm'' $L_i$ is the distance between the vertex of the 
track element and the point on the track element which is closest to OM $i$ 
and $\langle L \rangle$ is the average of all $L_i$-values.  
Track elements that fulfill these criteria 
qualify as \emph{track candidates} (TC).

Frequently, more than one track candidate is found.  In this case, a
cluster search is performed for all track candidates that fulfill the
quality criterion:
\begin{eqnarray}
\begin{array}{l}
\renewcommand{\arraystretch}{1.2}
Q_{\mbox{TC}}  \ge  0.7 \cdot Q_{\mbox{max}} \mbox{ ,~~~where}\\
Q_{\mbox{max}}  =  max(Q_{\mbox{TC}}) \mbox{~~~and} \\
\mbox{$Q_{\mbox{TC}}  =  min(N_{\mbox{AH}}, 0.3\,\mbox{m}^{-1} \cdot \sigma_L + 7)$}.
\renewcommand{\arraystretch}{1.0}
\end{array}
\end{eqnarray}
In the cluster search, the ``neighbors'' of each track candidate are
counted, where neighbors are track candidates with space angle
differences of less than $15^\circ$.  The cluster with the largest
number of track candidates is selected.

In the final step, the average direction of all track candidates
inside the cluster defines the initial track direction.  The track
vertex and time are taken from the central track candidate in the
cluster.  Well separated clusters can be used to identify independent
muon tracks in events which contain multiple muons (see
section~\ref{sec:bgclasses}).

\subsection{Line-Fit \label{sec:linefit}}

The
\emph{line-fit}~\cite{STENGER:1990:A} algorithm produces an
initial track on the basis of the hit times with an optional
amplitude weight.  It ignores the geometry of the \cer cone and
the  optical properties of the medium and assumes light traveling
with a velocity $\mathbf{v}$ along a 1-dimensional path through
the detector.  The locations of each PMT, $\mathbf{r}_i$,
which are hit at a time $t_i$ can be connected by a line:
\begin{equation}
\mathbf{r}_i\approx
\mathbf{r}+\mathbf{v} \cdot t_i~,
\label{eq:rdeflinefit}
\end{equation}
A $\chi^2$ to be minimized is defined as:
\begin{equation}
\chi^2\equiv \sum_{i=1}^{N_\mathrm{hit}}(\mathbf{r}_i-
\mathbf{r}-\mathbf{v} \cdot t_i)^2~,
\label{eq:chi2linefit}
\end{equation}
where $N_\mathrm{hit}$ is the number of hits.  
The $\chi^2$ is minimized by differentiation with respect to the
free fit parameters $\mathbf{r}$ and $\mathbf{v}$.
This can be solved analytically: 
\begin{equation}
\begin{array}{rclcrcl}
\mathbf{r}&=&\langle\mathbf{r}_i\rangle-
\mathbf{v}\cdot \langle t_i\rangle 
& \quad ~\mbox{and}~ \quad  & \mathbf{v}&=&
\frac{\displaystyle \langle\mathbf{r}_i\cdot t_i\rangle-
\langle\mathbf{r}_i\rangle\cdot \langle t_i\rangle}{\displaystyle \langle t_i^2\rangle-\langle t_i\rangle^2} ~,
\end{array}
\label{eq:rreslinefit}
\end{equation}
where 
$\langle x_i \rangle \equiv\frac{1}{N_\mathrm{hit}}\sum_i^{N_\mathrm{hit}}x_i$ 
denotes the mean of parameter $x$ with respect to all hits.

The line-fit thus yields a vertex point $\mathbf{r}$, and
a direction 
$\mathbf{e}=\mathbf{v}_\mathrm{LF}/|\mathbf{v}_\mathrm{LF}|$.
The zenith angle is given by
$\theta_\mathrm{LF}\equiv - 
\arccos (v_z/|\mathbf{v}_\mathrm{LF}|) $.

The time residuals (equation~\ref{eq:deftres}) for this initial track
generally do not follow the distribution expected for a \cer model.
If the $t_0$ parameter of the initial track is shifted to better agree
with a \cer model, subsequent reconstructions converge better (see
section~\ref{sec:coordtrafo}).

The absolute speed $v_\mathrm{LF} \equiv |\mathbf{v}|$, 
of the line-fit is the mean speed of the light
propagating through the 1-dimensional detector projection.
Spherical events (cascades) and high energy muons 
have low  $v_\mathrm{LF}$ values, and thin, long events 
(minimally ionizing muon tracks) have large values.

\subsection{Dipole Algorithm \label{sec:dipolefit}}

The \emph{dipole algorithm} considers the \emph{unit} vector from
one hit OM to the subsequently hit OM as an individual dipole moment.
Averaging over all individual dipole moments yields the global moment
$\mathbf{M}$.  It is calculated in two steps.  First, all hits are
sorted according to their hit times.  Then a \emph{dipole-moment}
$\mathbf{M}$ is calculated:
\begin{equation}
\mathbf{M} \equiv \frac{1}{N_\mathrm{ch}-1}\cdot
\sum_{i=2}^{N_\mathrm{ch}}
\frac{\mathbf{r}_i-\mathbf{r}_{i-1}}
{|\mathbf{r}_i-\mathbf{r}_{i-1}|}~.
\label{eq:dipolefit}
\end{equation}
It can be expressed via an absolute value $M_\mathrm{DA}\equiv|\mathbf{M}|$
and two angles $\theta_\mathrm{DA}$ and $\phi_\mathrm{DA}$.  These
angles  define the initial track.

The dipole algorithm does not generate as good an initial track as
the direct walk or the line-fit, but it is less vulnerable to a
specific class of background events: almost coincident atmospheric
muons from independent air showers in which the first muon hits the bottom 
and the second muon hits the top of the detector.

\subsection{Inertia Tensor Algorithm \label{sec:tensorfit}}

The \emph{inertia tensor algorithm} is based on a mechanical 
picture.  The pulse amplitude from a PMT at $\mathbf{r}_i$ 
corresponds to a virtual mass $a_i$ at $\mathbf{r}_i$. 
One can then define the  tensor of inertia $\mathbf{I}$ 
of that virtual mass distribution.
The origin 
is the center of gravity ($\mathbf{COG}$) 
of the mass distribution.  
The $\mathbf{COG}$-coordinates and the tensor of inertia components are given by: 
\begin{equation}
\begin{array}{lcl}
\renewcommand{\arraystretch}{2}
\mathbf{COG} &\equiv&  \sum\limits_{i=1}^{N_\mathrm{ch}} 
(a_i)^w\cdot 
\mathbf{r}_i \mbox{~~~and}\\
I^{k,l} &\equiv&   \sum\limits_{i=1}^{N_\mathrm{ch}} (a_i)^w \cdot 
[\delta^{kl}\cdot (\mathbf{r}_i)^2-r_i^k\cdot r_i^l]~.
 \label{eq:deftensorfit}
\renewcommand{\arraystretch}{1}
\end{array}
\end{equation}
The amplitude weight $w \ge 0$ can be chosen arbitrarily.  The
most common settings are $w = 0$ (ignoring the amplitudes) and 
$w = 1$ (setting the virtual masses equal to the amplitudes).
The tensor of inertia has three eigenvalues $I_j$, 
$j\epsilon\lbrace1,2,3\rbrace$, corresponding to its three main axes 
$\mathbf{e}_j$.  The smallest eigenvalue $I_1$ 
corresponds to the longest axis $\mathbf{e}_1$. 
In the  case of a long track-like event $I_1 \ll \{I_2, I_3\}$ and 
$\mathbf{e}_1$ approximates the direction of the track. 
The ambiguity in the direction along the $e_1$ axis is
resolved by choosing the direction where the average OM hit 
time is latest.  In the case of a cascade-like event, 
$I_1 \approx I_2 \approx I_3$.  The ratios between the $I_j$
can be used to determine the sphericity of the event.

\section{Aspects of the technical  Implementation \label{sec:technical}}

\subsection{Reconstruction Framework}

\begin{figure}[h!tbp]
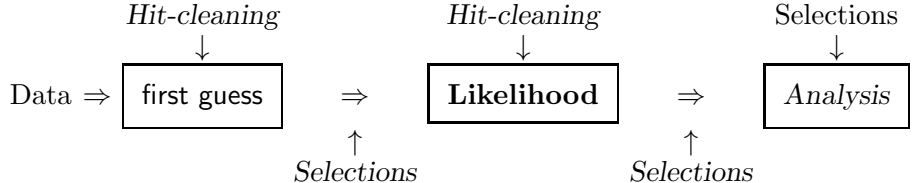

{ \small
$$
\renewcommand{\arraystretch}{0.85}
\begin{array}{rrccccc}
        &             &          \mbox{\sl{Hit-cleaning}}  &              &
        \mbox{\sl{Hit-cleaning}}   &            &  \mbox{Selections} 
        \\
        &             &          \downarrow  &              &
      \downarrow      &            &  \downarrow \\
\mbox{Data}  & \Rightarrow & \fbox{\sf{first guess}} & \Rightarrow & 
   \fbox{\bf{Likelihood}} & \Rightarrow & \fbox{\sl{Analysis}}\\
             &             &                        &\uparrow    &
                          & \uparrow     &    \\
             &             &                        &\mbox{\sl Selections} &
                          & \mbox{\sl Selections} &
\end{array}
\renewcommand{\arraystretch}{1}
$$
}
\caption[Analysis chain]{Schematic principle of the reconstruction chain
\label{fig:anachain}}
\end{figure}

The basic reconstruction procedure, sketched in figure~\ref{fig:anachain},
is sequential.  A fast reconstruction program calculates the initial track
hypothesis for the likelihood reconstruction.  All reconstruction programs
may use a reduced set of hits in order to suppress noise hits and other
detector artifacts.  Event selection criteria can be applied after each
step to reduce the event sample, and 
 allow more time consuming calculations at
later reconstruction stages.  This procedure may iterate with
more sophisticated but slower algorithms analyzing previous results.
The final step is usually the production of \emph{Data Summary Tape} 
(DST) like information, usually in
form of PAW N-tuples \cite{PAW:1999:A}.  A detailed description of this
procedure can be found in~\cite{JACOBSEN:1999:A}.

The reconstruction framework is implemented with the \emph{recoos}
program \cite{STREICHER//:2001:C}, which is based on the \emph{rdmc}
library~\cite{STREICHER//:2001:D} and the Si{EGM}u{ND} software
package ~\cite{STREICHER//:2001:B}.  The \emph{recoos} program is
highly modular, which allows flexibility in the choice and combination
of algorithms.

\subsection{Likelihood Maximization}

The aim of the reconstruction is to find the track hypothesis which
corresponds to the maximum likelihood. This is done by minimizing 
$ - \log (\mathcal{L}) $ with respect to the track parameters.  
We have implemented several
minimization procedures.

\begin{figure}[h!tb]
\centerline{\epsfig{file=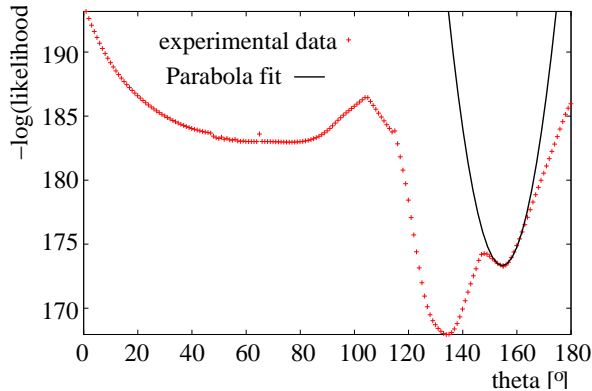,width=.49\textwidth}}
\caption[An example of the likelihood space]{An example 
of the likelihood space (1-dimensional projection) for a specific \aman event. 
Shown is  $ - \log (\mathcal{L}) $ as function of the zenith angle.
Each point represents a fit, for which  the zenith angle was fixed
and the other track parameters were allowed to vary in order to find the 
best minimum. 
A local minimum which was found by a gradient likelihood minimization is 
indicated by a fitted parabola. Improved methods that avoid this are 
described in the text.
\label{fig:likespace}
}
\end{figure}

The likelihood space for \aman events is often  characterized 
by several minima.  
Local likelihood minima can
arise due to symmetries in the detector, especially in the azimuth
angle, or due to unexpected hit times caused by scattering.  In the
example, shown in figure~\ref{fig:likespace}, the reconstruction
converged on a local minimum, because of non-optimal starting values.
Several techniques, which are used to find the global minimum, are
here presented.  In particular, the iterative
reconstruction, section~\ref{sec:iterreco}, solves the problem and
converges to the global minimum.
One generally assumes that the global minimum corresponds to the true
solution, but this is not always correct
 due to stochastic nature of light emission and detection.
Such events cannot be reconstructed properly and have to be rejected
using quality parameters (see section \ref{sec:parameters}).

\subsubsection{Minimization algorithms}

The reconstruction framework allows us to use and compare
these numerical minimization algorithms:
\emph{Simplex}~\cite{PRESS/TE/VE/FL:1997:A},
\emph{Powell's}~\cite{PRESS/TE/VE/FL:1997:A},
\emph{Minuit}~\cite{JAMES:1994:A} (using the \emph{minimize} method),
and \emph{Simulated annealing}~\cite{PRESS/TE/VE/FL:1997:A}.
The Simplex algorithm is the fastest algorithm.
Powell's method and Minuit are
$\tld 5$~times slower than the Simplex algorithm. 
 The reconstruction
results from Minuit and the Simplex algorithm are nearly identical and
almost as good as the Powell results.  Exceptions occur in less than
1\% of the cases, when these methods fail and stop at the extreme 
zenith angles
$\theta=0^\circ $ and $\theta=180^\circ $.  The Simulated
annealing algorithm is less sensitive to local minima than the other
algorithms, but it is much slower and  requires fine-tuning. 

\subsubsection{Iterative Reconstruction\label{sec:iterreco}}

The \emph{iterative reconstruction} algorithm successfully copes with
the problem of local minima and  extreme zenith angles by performing
multiple reconstructions of the same event.  Each reconstruction starts
with a different initial track.  Therefore, the fast Simplex algorithm
is sufficient.

The ability to find the global minimum depends strongly on the quality
of the initial track.
A systematic scan 
 of the full parameter space
for initial seeds
is not feasible.  Instead
the iterative algorithm concentrates on the direction angles, zenith
and azimuth, and uses reasonable values for the spatial
coordinates.  The following procedure yields good results.

The result of a  first minimization
is saved as a reference.  
Then both direction angles are randomly selected.  The track point,
$\mathbf{r_0}$, is transformed to the point on the new track, which is
closest to the center of gravity of hits.  The time, $t_0$, of this
point is shifted to match the \cer expectation (see
section~\ref{sec:coordtrafo}).  
Then a new minimization is started.
If the minimum is less than the
reference minimum, it is saved as the new reference.  This procedure
is iterated $n$ times,
and the best minima found for zenith angles \emph{above} and
\emph{below} the horizon, are saved, and used to generate an important
selection parameter (see section~\ref{sec:parameters}).

This algorithm substantially reduces the number of false
minima found, after a few iterations.  For $n=6$ roughly $ 95 $\% of the
results are in the vicinity of the asymptotic optimum for $n \to
\infty$.  For $n \simeq 20 $ more than $99$\% of the results are the
global minimum.
Despite  the fast convergence, the \emph{iterative reconstruction}
requires significant CPU time, which limits its use to reduced data sets.

\subsubsection{Lateral shift and time residual  \label{sec:coordtrafo}}

The efficiency of finding the global minimum of the
likelihood function can be improved by translating the arbitrary vertex
and/or time origin of the output track from the first guess algorithm
before application of the full maximum likelihood method.

\emph{Transformation of~$\mathbf{r_0} $}.
In general this ``vertex'' point is arbitrary in the infinite
track approximation used, and first guess methods
may produce positions distant from the detector. During the likelihood
minimization, numerical errors can be avoided and the convergence improved
by shifting this point along the direction of the track towards 
the point closest to the \emph{center of gravity of hits} 
(see equation~\ref{eq:deftensorfit}). 
The vertex time $t_0$ is 
transformed accordingly: $\Delta(t_0)=\Delta(\mathbf{r_0})/c$.

\emph{Transformation of~$t_0 $}.
The time $t_0$ obtained from first guess algorithms is
not calculated from  a full \cer model. The efficiency of the likelihood
reconstruction can be improved by shifting the $t_0$
 such that the time residuals, equation~\ref{eq:deftres},
fit better to a \cer model.
In particular it is useful to avoid negative
$t_\mathrm{res} $, which would correspond to causality violations.
This can be achieved by transforming   $t_0 \to t_0 - t_\mathrm{res}^-$, where
$ t_\mathrm{res}^-$ is the most negative time residual.

\subsubsection{Coordinates and restricted parameters \label{sec:penalty}}

The track coordinates  $\mathbf{a}$, which are used by the likelihood,
are independent of the  coordinates
actually chosen for the minimization.  Therefore,
the coordinate system can be chosen arbitrarily.
Any of the parameters in this coordinate system can be kept fixed.
During the minimization \emph{parameterization}
functions translate the coordinates as necessary.
The most commonly used coordinates are
$\mathbf{r_0}$ and the zenith and azimuth angles $\theta $, $\phi $.

The freedom in the choice of
coordinates  can be used to improve the numerical minimization,
for systematic studies, or to fix certain parameters  according to 
external knowledge.
An example is the reconstruction of coincident events with 
the \spase surface arrays \cite{AHRENS//:2002:E}.
Here, we fix the location of the trajectory to coincide with the core location at the surface as measured by \spase. Then, the direction is determined with the \aman reconstruction subject to this constraint.

Under certain circumstances the
allowed range of the reconstruction parameters is restricted.
The most important example here is to restrict the reconstructed 
zenith angle to above or below the horizon, to find the most likely
up- or down-going tracks, respectively.
Comparing the quality of the two solutions can be used for 
background rejection.
Technically the constrained fit is accomplished 
by multiplying the  likelihood by a prior,
which is zero outside the allowed parameter range.

\subsection{Preprocessing and Hit cleaning\label{sec:prepro}}

The data must be filtered and calibrated before reconstruction.
Defective OMs are removed, and the amplitudes and hit times are
calibrated.  A \emph{hit cleaning} procedure identifies and flags
hits which appear to be noise or electronic effects, such as cross
talk or after-pulsing.  These hits are not used in the reconstruction,
but they are retained for post-reconstruction analysis.

The hit cleaning procedure can be based on simple and robust
algorithms, because the 
PMTs have low noise rates.  Noise and
after-pulse hits are strongly suppressed by rejecting hits that are
isolated in time and space from other signals in the detector.
Typically a hit is considered to be noise if there is no hit within a
distance of $60\,$m to $100\,$m and a time of $\pm 300\,$ns to $\pm
600\,$ns.  Cross talk hits are identified by examining the amplitudes
and pulse widths of the individual pulses and by analyzing the
correlations of uncalibrated hit times with hits of large amplitude
in channel combinations which are known to cross talk to each other.
The required cross talk correlation map was determined independently
in a dedicated calibration campaign.

\subsection{Processing Speeds \label{sec:speed}}

The first guess algorithms are sufficiently fast that the execution
time is dominated by file input/output and the software framework.  Typical fit
times are $\approx 20\,$ms per event on a 850\,MHz
Pentium-III Linux PC.  The processing speed of the
likelihood reconstructions can vary significantly depending on the
number of free parameters, the number of iterations, the
minimization algorithm, and the experimental parameters like the number 
of hit OMs.  These effects dominate the differences in
processing speeds due to different reconstruction algorithms.  The
typical execution time for a 16-fold iterative likelihood
reconstructions using the simplex minimizer to reconstruct the 5~free
track parameters is $\simeq 250\,$ms per event.

\section{Background Rejection \label{sec:reject}}

The performance of the reconstruction depends strongly on the quality
and background selection criteria.  The major classes of background
events in \aman (see section~\ref{sec:bgclasses}) are suppressed by
the quality parameters presented in section~\ref{sec:parameters}.
Optimization strategies for the selection criteria are summarized in
section~\ref{sec:strategy}.  Finally, we evaluate the reconstruction
performance in section~\ref{sec:perform}.

\subsection{Background Classes \label{sec:bgclasses}}

Most background events from atmospheric muons are well reconstructed
and can be rejected by selecting up-going reconstructed events.
However, there is a small fraction of mis-reconstructed
events, amounting to about $ 10^{-2}$ for the unbiased
and  about $ 10^{-4}$ for the zenith-weighted reconstruction.
These events are rejected by additional selection
criteria described in section~\ref{sec:parameters}.  These background
events are classified as follows.

\begin{description}

\item[Nearly horizontal muons:] These events have true incident angles
close to the horizon.  A small error in the reconstruction causes them
to appear as up-going.  These events are not severely
mis-reconstructed, but occur due to the finite angular resolution.

\item[Muon bundles:] The spatial separation between multiple muons
from a single air shower, a \emph{muon bundle}, is usually small
enough that the event can be described by a single bright muon track.
If the separation is too large, the reconstruction fails.

\item[Cascades:] Bright stochastic energy losses (e.g.\ 
bremsstrahlung) produce additional light, which distorts the \cer
cone from the muon.  Cascades emit most of their light with the same
\cer angle as the muon, but some light is emitted at other
angles.  These secondary events can cause the reconstruction to fail,
especially when the cascade(s) produce more light than the muon
itself.  A special class of these events are muons which pass outside
the detector and release a bright cascade, which can mimic an up-going
hit pattern.

\item[Stopping muons:] Over the depth of the detector the muon flux
changes by a factor of $\tld 2$, since muons lose their energy and
stop.  These muons can create an up-going hit pattern, especially when
the muon stops just before entering the detector from the side.

\item[Scattering layers:] 
The  scattering of light in the polar ice cap varies
with depth.
Light from bright events, 
can mimic an up-going hit pattern, 
in particular when it  traverses layers of higher scattering.

\item[Corner clippers:] These are events where the muon passes
diagonally below the detector.  The light travel upwards through the
detector mimicking an up-going muon.

\item[Uncorrelated coincident muons:] Due to the large size of the
\aman detector, the probability of muons from two independent air
showers forming a single event is small on the trigger level but not
negligible.  If an initial muon traverses the bottom of the detector
and a later muon traverses the top, the combination can be reconstructed as
an up-going muon.

\item[Electronic artifacts:] Noise, cross talk and other transient
electronic malfunctions are generally small effects, but they can
occasionally produce hits, which distort the time pattern.  
Such effects become important after a selection process of several orders
of magnitude.
\end{description}

\subsection{Quality Parameters \label{sec:parameters}}

Background events, which pass a zenith angle selection, need to be
rejected by applying selection criteria on quality parameters.  These
parameters usually evaluate information, which is not optimally
exploited in the reconstruction.  The detailed choice of quality
parameters is specific to each analysis.  Here, we summarize the most
important categories.

The \emph{number of direct hits}, $N_\mathrm{dir}(t_1:t_2)$, is the
number of hits with small time residuals: $ t_1 < t_\mathrm{res} <t_2
$ (see equation~\ref{eq:deftres}).  Un-scattered photons provide the best
information for the reconstruction, and
a large number of $N_\mathrm{dir}$  
indicates high quality information in the event.
  Empirically reasonable values are
$t_1 \simeq -15\,$ns and $t_2 $ between $=+25\,$ns and $+150\,$ns,
depending on the specific analysis.

The \emph{length of the event} $L$ is obtained by projecting each hit OM 
onto the reconstructed track and taking the distance between the two outermost
of these points.  $L$ can be considered as the ``lever arm'' of the
reconstruction.  Larger values corresponding to
a more robust and precise reconstruction of the track's direction.
This parameter is particularly powerful when calculated
for direct hits only, and is then referred to as $L_\mathrm{dir}(t_1:t_2)$.
Length requirements are  efficient against corner clippers, stopping muons 
and cascades. 

The absolute value of the likelihood at the maximum is a good
parameter to evaluate the quality of a reconstruction.  Here, a useful
observable is the \emph{likelihood parameter} $\mathsf{L}$ which is
defined as
\begin{equation}
\mathsf{L}\equiv - \frac{\log(\mathcal L)}{N_\mathrm{free}}~,
\label{eq:deflikepara}
\end{equation}
where $N_\mathrm{free} $ is the degrees of freedom
(e.g.~$N_\mathrm{free} = N_\mathrm{hits}-5$ for a track
reconstruction).  For Gaussian probability distributions this
expression corresponds to the reduced chi-square.
$\mathsf{L}$ can be used as a selection parameter, smaller values
corresponding to higher quality. 
A selection of events with good
$\mathsf{L}^{P^\mathrm{hit}P^\mathrm{no-hit}}$ values is efficient
against stopping muons.

Comparing $\mathsf{L}$ from different reconstructions is a
powerful technique.  Cascade-like events will have a better
likelihood from a cascade reconstruction than one from a 
track reconstruction.

Another efficient rejection method is to compare $\mathsf{L}$ for the
best up-going versus the best down-going reconstruction of a single
event.  If the up-going reconstruction is not significantly better
than the down-going reconstruction, the event is rejected.  These
values can be obtained from the iterative reconstruction method
(section~\ref{sec:iterreco}) or by restricting the parameter space.
This method is particularly powerful when the down-going
reconstruction uses a zenith weighted likelihood 
(section \ref{sec:weightedreco}).

The reconstruction methods consider the p.d.f.\ for each hit separately
but ignore correlations.  Therefore, the reconstructions assign the
same likelihood to tracks where all hits cluster at one end of the
reconstructed track and tracks where the same number of hits are
smoothly distributed along the track.  The latter hit pattern
indicates a successful track reconstruction, while the former hit
pattern may be caused by a background event.  The \emph{smoothness}
parameter $S$ was inspired by the Kolmogorov-Smirnov test of the
consistency of two distributions.  $S$ is a measure of the consistency
of the observed hit pattern with the hypothesis of constant light
emission by a muon.  The simplest definition of the smoothness $S$ is
$S = S_j^\mathrm{max}$, where $S_j^\mathrm{max}$ is that $S_j$, which
has the largest absolute value, and $S_j$ is defined as
\begin{equation}
\renewcommand{\arraystretch}{1.5}
S_j \equiv \frac{\displaystyle j -1} {\displaystyle N-1} - 
        \frac{\displaystyle l_j}{\displaystyle l_N}
\renewcommand{\arraystretch}{1}
~.
\label{wq:defsmooth}
\end{equation}
$l_j$ is the distance along the track between the points of closest
approach of the track to the first and the $j^\mathrm{th}$ hit module,
with the hits taken in order of their projected position on the track.
$N$ is the total number of hits.  Tracks with hits clustered at the
beginning or end of the track have $S$ approaching $+1$ or $-1$,
respectively.  High quality tracks with $S$ close to zero, have hits
equally spaced along the track.  A graphical representation of the
smoothness construction can be found in~\cite{AHRENS//:2002:C}.

Extensions of this smoothness parameter include the restriction of the
calculation to direct hits only or using the distribution of hit times
$t_i$ instead of the distances $l_i$.

A particularly important extension is $S^{P^\mathrm{hit}}$.  In order
to account for the granularity and asymmetric geometry of the detector
one can replace the above formulation with one that models the hit
smoothness expectation for the actual geometry of the assumed muon
track.  This can be accomplished by using the hit probabilities of all
$N_\mathrm{OM}$, the number of operational OMs, (ordered along the track)
as weights: $S^{P^\mathrm{hit}} =
\max{(S_j^{P^\mathrm{hit}})}$ with
\begin{equation}
\renewcommand{\arraystretch}{1.5}
S_j^{P^\mathrm{hit}} \equiv 
\frac{\sum_{i=1}^j \Lambda_i}{\sum_{i=1}^{N_\mathrm{OM}} \Lambda_i } - 
        \frac{\sum_{i=1}^j P^\mathrm{hit,i} }{\sum_{i=1}^{N_\mathrm{OM}} 
          P^\mathrm{hit,i} }
\renewcommand{\arraystretch}{1}
~.
\label{wq:defphitsmooth}
\end{equation}
$\Lambda_i =1 $, if the OM $i$ was hit and 0 otherwise and $
P_\mathrm{hit,i} $ is the probability for OM $i$ to be hit given the
reconstructed track.  The hit probabilities are calculated according
to the algorithm in section~\ref{sec:phitreco}.  Smoothness selections
are very efficient against secondary cascades, stopping muons and
coincident muons from independent air showers.

Interesting \aman events are analyzed with multiple reconstruction
algorithms.  An event is most likely to have been reconstructed
correctly, if the different algorithms produce consistent results.

For two reconstructions with directions $\mathbf{\mathbf{e}}_1$ 
and $\mathbf{\mathbf{e}}_2 $, the space angle between them is given by 
$\Psi =  \arccos{ (\mathbf{\mathbf{e}}_1 \cdot \mathbf{\mathbf{e}}_2 )}$,
which should be reasonably small for successful reconstructions.
This concept can be extended to multiple reconstructions and
their angular deviations  from the  average direction.
For $n$ different reconstructed directions, $\mathbf{\mathbf{e}}_i$, 
the average reconstructed direction, $\mathbf{\mathbf{E}}$, is given by 
$\mathbf{\mathbf{E}}=\sum_i^n{\mathbf{\mathbf{e}}_i}/
|\sum_i^n{\mathbf{\mathbf{e}}_i}|$.
We can define the parameter
\begin{equation}
\Psi_w=\left(\sum_i [ \arccos{(\mathbf{\mathbf{e}}_i \cdot \mathbf{\mathbf{E}})} ]
^w\right)^{1/w}~.
\label{eq:defpsi}
\end{equation}
$\Psi_1$ describes the average space angle between the individual
reconstructions and $\mathbf{\mathbf{E}}$.  $\Psi_2$ is a different
parameter, which treats the deviations between
$\mathbf{\mathbf{E}}$ and the $\mathbf{\mathbf{e}}_i$ as ``errors''
and adds them quadratically.  Small values of $\Psi_1$
or $\Psi_2$ indicate consistent reconstruction results.

The $\Psi$ parameters are a mathematically correct consistency check
only when comparing the results of uncorrelated reconstructions of the
same intrinsic resolutions.  This is not the case when comparing
different \aman reconstructions.  Irrespective of the validity of such
an interpretation, $\Psi_1$ or $\Psi_2$ are very efficient selection
criteria, especially against almost horizontal muons and wide muon
bundles.

A few  additional selection parameters are closely related to first guess
methods. 
The ratio of the eigenvalues of the \emph{tensor of inertia}
(see section~\ref{sec:tensorfit}) are a measure of the sphericity of
the event topology, which is an efficient selection parameter against
cascade backgrounds.  Tracks reconstructed as down-going by the
\emph{dipole fit} (see section~\ref{sec:dipolefit}) that have a
non-negligible \emph{dipole moment}, $M_{DA} \equiv |\vec{M}|$,
indicate coincident muons from independent air showers.
Larger  values of the line-fit speed $v_\mathrm{LF}$ 
(see section~\ref{sec:linefit}) are an indication for longer
muon-like, smaller values  for more spherical cascade-like
events.

Finally, two  approaches evaluate the ``intrinsic
resolution'' or ``stability'' of the reconstruction of each event.
One approach quantifies the sharpness of the minimum found by the
minimizers in $-\log(\mathcal L)$ by fitting a paraboloid to it.  The
fitted parameters can then be used to classify the sharpness of the
minimum.  The other approach splits an event into sub-events
(for example, containing odd- vs. even-numbered hits)
 and
reconstructs the sub-events.  If the reconstructed directions of the
sub-events are different, then the reconstruction of the full event
has a larger uncertainty.

\subsection{Analysis Strategies\label{sec:strategy}}

Analyses that search for neutrino induced muons must  cope with
a large background of atmospheric muons.  The optimal choice of
reconstruction and selection criteria varies strongly with different
expectations for the energy and angular distribution of the signal
events.  The goal is to optimize the signal efficiency over the  
background  or noise (square root of the background)
 based on sets of signal and background data.

\begin{itemize}

\item The selection criteria for background sensitive variables may be
adjusted individually such that a specified fraction of signal events
pass.  After these \emph{first level criteria} are set, the adjustment
is repeated until the desired background rejection is reached.  Each
iteration defines a ``cut-level'', which corresponds to data sets of
increasing purity.  This simple method 
is used  to derive a defined set of selection parameters for the
performance section~\ref{sec:perform}.  However, less efficient
criteria are mixed with more efficient criteria, and correlations of
the variables are not taken into account.  Therefore, this method does
not achieve the optimum signal efficiency.

\item An improvement to this method has been demonstrated in an \aman
point source analysis \cite{YOUNG:2001:A,AHRENS//:2002:F}.
Here, a selection criterion is only applied to the most sensitive variable,
and the most sensitive variable is  determined at each cut level.
An interesting aspect of this point source search is that the
experimental data themselves can be used as a background sample, which
reduces systematic uncertainties from the background simulation.  The
selection criteria are not optimized with respect to signal purity but
with respect to an optimal significance of a possible signal.

\item Another approach is to combine the selection parameters into a
single selection parameter, called \emph{event quality}.  This can be
done by rescaling and normalizing each of the selection parameters
according to the cumulative distribution of the signal expectation.
The \aman analyses of atmospheric neutrinos
\cite{AHRENS//:2002:C,BOESER:2002:A}  used this technique.

\item 
Additional approaches use discriminant analysis\cite{AHRENS//:2002:B} or
neural nets\cite{HUNDERTMARK//:2001:A,NIESSEN//:2001:A,BIRON:1998:A} to
optimize the efficiency while taking into account the correlations between
selection criteria and their individual selectivity.  However, both methods
depend critically on a good agreement between experiment and simulation.
These methods quantify the efficiency of each parameter by including
and excluding it from the optimization procedure.

\item 
The ``\cuteval'' method finds the optimum combination of selection
parameters and cut values by numerically maximizing a significance function,
$\mathcal{Q}$.  An example is $\mathcal{Q} = S/\sqrt{B}$, where $S$ is
the number of signal events, and $B$ is the number of background events
after selection.

The implementation proceeds in several steps.  First, the most efficient
selection parameter, $\mathcal{C}_1$, is the parameter that individually
maximizes $\mathcal{Q}$.  The next parameter, $\mathcal{C}_2$, is the
parameter that maximizes $\mathcal{Q}$ in conjunction with $\mathcal{C}_1$.
More parameters are successively determined until the addition of a new
parameter fails to improve $\mathcal{Q}$.  This procedure takes
correlations between the selection criteria into account.  The final number
of selection parameters is reduced to a minimum, while maximizing the
efficiency.
Next, the optimal selection for this combination parameters is computed 
as a function
of a boundary condition (e.g. the maximum number of
accepted background events).
This boundary condition is also  used to define 
a  single quality parameter.

Such a formalized procedure  has to be carefully monitored, e.g. to 
handle potentially un-simulated experimental effects.
The \cuteval  procedure is monitored
by defining different, complementary optimization functions $\mathcal{Q} $,
which allow real and simulated data to be compared
\cite{AHRENS//:2002:C,GAUG:2001:A,GAUG:2000:A,GAUG:2000:B,LEUTHOLD:2001:A}.

\end{itemize}

\section{Performance \label{sec:perform}}

This section describes the  performance of the 
 reconstruction methods.
It is based on illustrative  data selections, and
the actual performance of a dedicated analysis
can be  different.
Unless noted otherwise, the data shown is from Monte Carlo simulations of
atmospheric neutrinos for \amaii.

\subsection{First Guess Algorithms \label{sec:firstguessproc}}

Since the first guess algorithms are used as
a starting point for the full reconstruction, they should 
provide a reasonable estimate
of the track coordinates. Also, these algorithms are used
as the basis of early level filtering, and therefore need to be
sufficiently accurate for that purpose, i.e. they should at least 
reconstruct the events in the correct hemisphere.

\begin{table}[!hbpt]
\begin{center}
\begin{tabular}{|l|c|c|c|} 
\hline
reconstruction & atm. $\mu$ & atm. $\nu$ \\
\hline
direct walk                     &  1.5\% & 93\% \\
line-fit                        &  4.8\% & 85\% \\
dipole algorithm            & 16.8\% & 78\% \\
\hline
\end{tabular}
\end{center}
\caption[Background Suppression of first guess algorithms]{The 
atmospheric muon and atmospheric neutrino detection efficiencies for a 
selection at $\theta \ge 80^\circ$ for the first guess algorithms.
\label{tab:bgsup}
}
\end{table}

As an example, Table~\ref{tab:bgsup} gives the passing efficiencies
with respect to the \amaii trigger
for atmospheric neutrinos (signal) and atmospheric muons (background)
for the first guess methods (see section~\ref{sec:firstguess}),
after the  selection of events with calculated zenith angles 
larger $80^\circ$.
 The \emph{direct walk} algorithm gives the
best background suppression and the highest atmospheric neutrino passing
rate.  Correspondingly, it also gives the best initial tracks to the likelihood
reconstructions.

\subsection{Pointing accuracy  of the Track Reconstruction\label{sec:reso}}

The angular accuracy of the reconstruction can be expressed in terms of
a point spread function, which is given by  the space angle
deviation  $\Psi$ between the true  and the reconstructed
direction of a muon corrected for solid angle. 
The space angle deviation is a combined result
of two effects: a systematic shift in the direction 
and a random spread around this shift.
 In a point source analysis, for example, it is possible to correct for
systematic shifts and be limited by the point spread function alone
\cite{AHRENS//:2002:F}.

\begin{figure}[h!tb]
\centerline{\epsfig{file=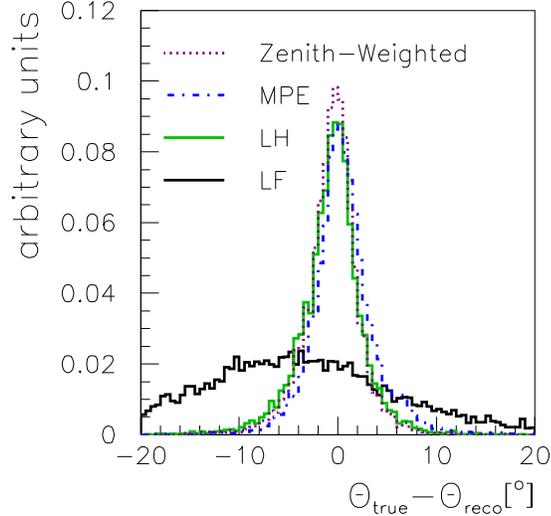,width=.45\textwidth}}
\caption[The dependence of the zenith angle
deviation on the reconstruction algorithm]
{The zenith angle deviations for various reconstructions of \amax.  
The result of an atmospheric neutrino simulation 
after the selection criteria of \cite{AHRENS//:2002:C} is shown.
The fits are  a \emph{line-fit} (LF), an iterated \emph{upandel} fit (LH), 
an iterated zenith-weighted \emph{upandel} fit and a MPE fit.
\label{fig:zen_resolution}
}
\end{figure}

\begin{figure}[h!tb]
\centerline{\epsfig{file=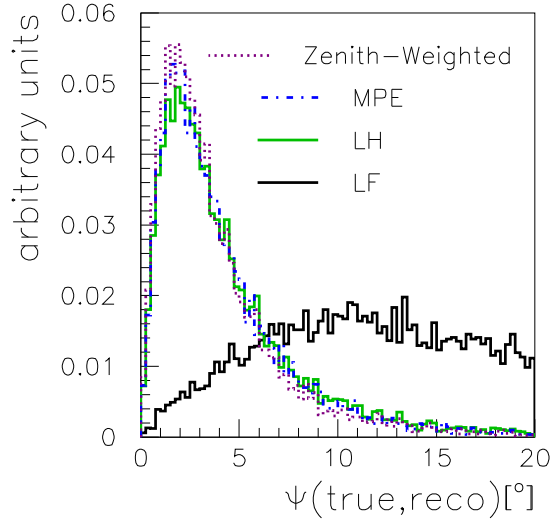,width=.45\textwidth}}
\caption[The dependence of the space angle deviation
on the reconstruction algorithm]
{The distribution of space angle deviations 
for various reconstructions of \amax.  
The result of an atmospheric neutrino simulation 
after the selection criteria of \cite{AHRENS//:2002:C} is shown.
The fits are  a \emph{line-fit} (LF), an iterated \emph{upandel} fit (LH), 
an iterated zenith-weighted \emph{upandel} fit and a MPE fit.
\label{fig:psi_resolution}
}
\end{figure}

The zenith and space angular deviations are shown in
figures~\ref{fig:zen_resolution} and \ref{fig:psi_resolution}. 
They are obtained
by the reconstruction algorithms as used in \amax.  The same event
selection is used for all. As a general observation, 
the distributions of deviations for different reconstruction
algorithms is surprisingly similar after a particular selection.  Larger 
 differences are usually seen in the selection efficiencies.  
A similar behavior is observed  for \amaii.

\begin{figure}[h!tb]
\centerline{\epsfig{file=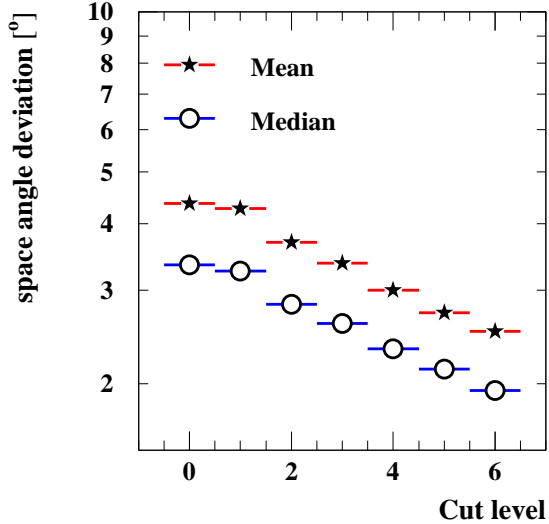,width=.45\textwidth}}
\caption[The dependence of the space angle deviation on the cut level]
{The dependence of the space angle deviation of the LH reconstruction
in \amaii  on the event selection (cut levels).
\label{fig:res_lev}
}
\end{figure}

\begin{figure}[h!tb]
\centerline{\epsfig{file=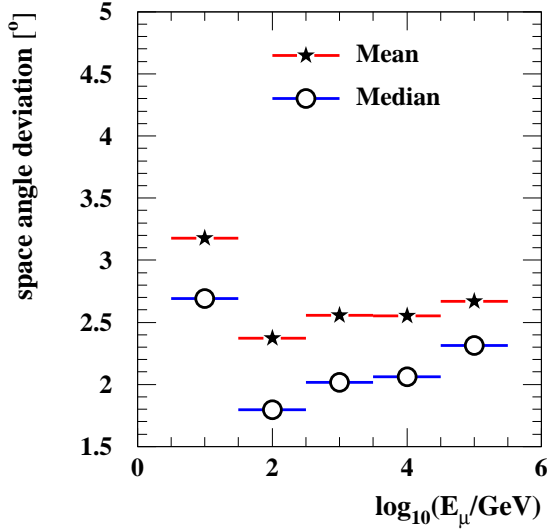,width=.45\textwidth}}
\caption[The dependence of the resolution on $E_\mu$]
{The dependence of the  space angle deviation of the LH fit on 
the muon  energy for \amaii. Shown are mean (stars) 
and median (circles)  for  simulated atmospheric neutrinos.
\label{fig:res_spec_energy}
}
\end{figure}

\begin{figure}[h!tb]
\centerline{\epsfig{file=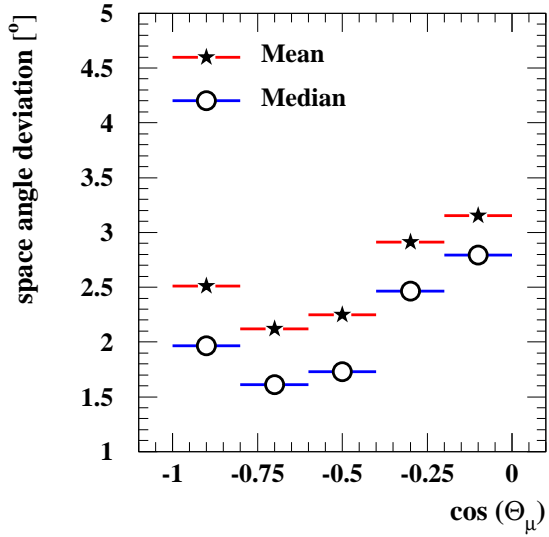,width=.45\textwidth}}
\caption[The dependence of the space angle deviations
on $\cos{\theta_\mu}$]
{The space angle deviations
of the LH fit as a function of the cosine of the incident
zenith angle (for \amaii). Shown are the
mean (stars) and median (circles) for  simulated atmospheric neutrinos.
\label{fig:res_angle}
}
\end{figure}

\begin{figure}[h!tb]
\centerline{\epsfig{file=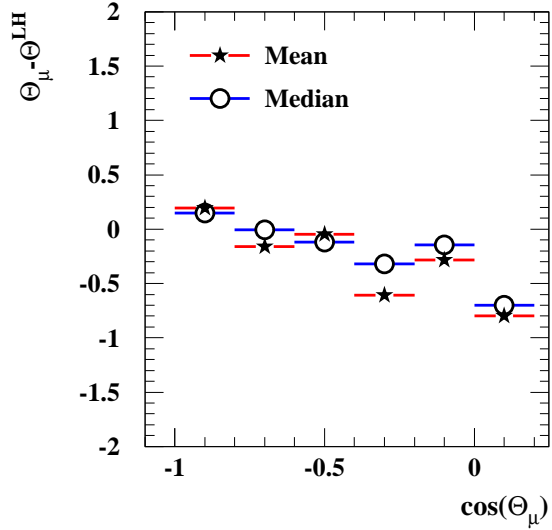,width=.45\textwidth}}
\caption[The zenith angle offset]%
{The zenith angle shift of the reconstruction versus the cosine of the
  incident  angle.
Shown are  mean (stars) and median (circles) for
simulated atmospheric neutrinos.
\label{fig:zen_shift}
}
\end{figure}

The dependence of the  space angle deviation for the full \amaii detector
on the cut
level\footnote{
The cut levels defined here are typical and intended
as demonstrating  example. 
We use typical selection parameters  
from section \ref{sec:parameters}:
 the reconstructed zenith angle,
$\theta^\mathrm{DW}>80^\circ$, $\theta^\mathrm{LH}>80^\circ$,
 $N_\mathrm{ch}$,
 $N_\mathrm{dir}^\mathrm{LH}(-15:25)$,
 $L_\mathrm{dir}^\mathrm{LH}(-15:75) $,
 $\mathsf{L}^\mathrm{LH}$,
 $S^\mathrm{LH}$ and
 $\Psi_1(DW,LH,MPE)$.
Our goal here is to illustrate the analysis, and we do not optimize with
respect to efficiency and angular resolution.  Instead each individual
criterion is enforced in such a way that 95\% of the events from the
previous level would pass, and correlations between the parameters are
ignored.  Specific physics analyses will use selection criteria of higher
efficiency and will achieve better angular resolutions than the $\simeq
2^\circ$, shown here.
} 
for the LH reconstruction is shown in figure~\ref{fig:res_lev}.   
The tighter the
selection criteria, the better the angular resolution. The same
general trend is true for the other reconstructions.
  Tight criteria
select events with unambiguous hit topologies, which are reconstructed better.
The results for cut level~6 are shown in
figures~\ref{fig:res_spec_energy}-\ref{fig:zen_shift} as function of the 
energy and the zenith angle.

The angular resolution (see figure~\ref{fig:res_spec_energy}) has a
weak energy dependence.  The energy of the muon is taken at the point of
its closest approach to the detector center.  Best results are achieved for
energies of $100\,$GeV to $10\,$TeV.  At energies $<100\,$GeV, the muons
have paths shorter than the full detector, which limits the angular
resolution.  At energies $>10\,$TeV, more light is emitted due to
individual stochastic energy loss processes along the muon track.  Here, the
hit pattern is not correctly described by the underlying reconstruction
assumption of a bare muon track (see section~\ref{sec:bgclasses}).

The space angular resolution depends
on the incident muon
zenith angle (see figure~\ref{fig:res_angle}).  Again this is shown only
for the LH reconstruction, the other reconstructions are similar.   
Up-going muons with  $\cos{\theta_\mu}\simeq -0.7$
are best reconstructed, and horizontal muons are
the worst, because of the geometry of the \amaii detector. 
 Nearly vertical
events with $\cos{\theta_\mu}\simeq -1$ have a poorer angular resolution,
because they illuminate fewer strings, which can cause ambiguities in the
azimuth.

Systematic shifts also degrade the angular resolution.  \aman observes a
small zenith dependent shift of the reconstructed zenith angle and no
systematic shift in azimuth.  This is shown in figure~\ref{fig:zen_shift}
for simulated atmospheric neutrinos in \amaii.  The size of this shift
depends on the zenith angle itself, and it is determined by the geometry of
\aman, which has a larger size in vertical than in horizontal directions.
From a comparison with \amax data~\cite{BIRON:2002:A,YOUNG:2001:A}, we
observe that these shifts become smaller with a larger horizontal detector
size.  These shifts are confirmed by analyzing \aman events  coincident
with \spase (see below).

\begin{figure}[h!tb]
\centerline{\epsfig{file=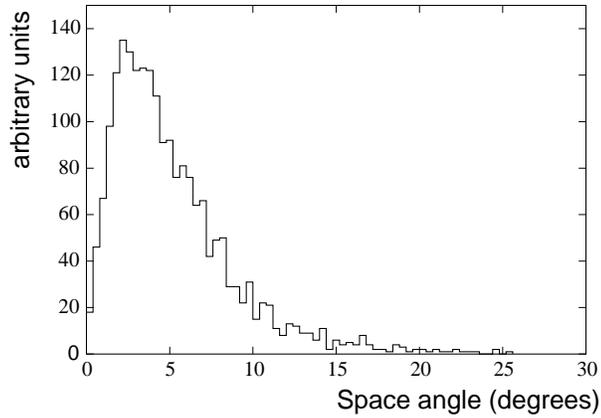,width=.49\textwidth}}
\caption[Difference between \space and \amax reconstructions]
{Distribution of the space angle deviations between air shower directions 
assigned by \spaseii and muon directions assigned by \amax for 
coincident events measured in 1997. The figure is not corrected for
the systematic shift.
\label{fig:spase}
}
\end{figure}

These angular deviations  have been obtained from Monte Carlo 
simulations.  They
can be experimentally verified by analyzing coincident events between \aman
and \spase. An analysis of data from the 10~string \amax detector, shown in
figure~\ref{fig:spase}, confirms the estimate of $\simeq 3^\circ $ obtained
from Monte Carlo studies for
\amax~.  Unfolding the
estimated \spase resolution of $\simeq 1^\circ $ confirms the estimated \amax
resolution of $\simeq 3^\circ$ near the \spase-\aman coincidence
direction \cite{AHRENS//:2002:E,BAI//:2001:A,RAWLINS:2001:A}.

A simulation-independent estimate can be obtained by splitting the hits of
individual events in two parts and reconstructing each sub-event
separately.  The difference in the two results gives an estimate of the
total angular resolution.  Such analyses are being performed at present and
results will be published separately.

\subsection{Energy Reconstruction \label{sec:energyperform}}

\begin{figure}[htpb]
\centerline{
\epsfig{file=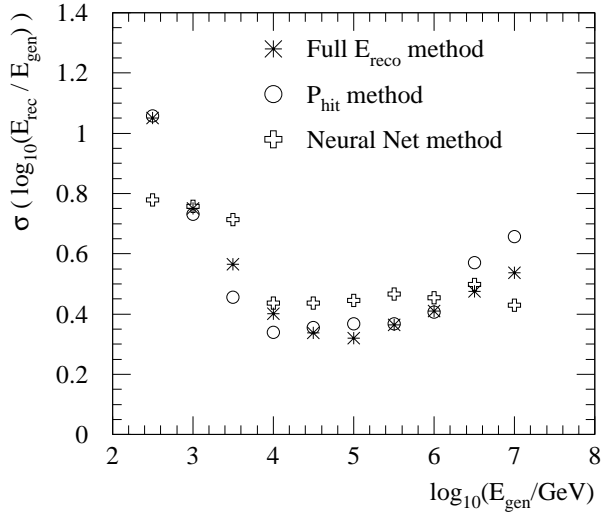,width=0.49\textwidth}
}
\caption{Comparison of the resolution of three different 
energy reconstruction approaches for \amax.
$E_\mathrm{gen}$ is the generated energy (MC) and $E_\mathrm{rec} $
the reconstructed energy.
\label{fig:energyres}}
\end{figure}

The energy resolution of the three methods, described in
section~\ref{sec:energyreco}, is shown in figure~\ref{fig:energyres} as
function of the muon energy at its closest point to the  \amax center.  
The resolution for
\amax in $\Delta \log_{10} E $ is $ \simeq 0.4$, for the interesting energy range
of a few TeV to $1\,$PeV.  Below $\simeq 600$\,GeV the energy resolution is
limited, because the amount of light emitted by a muon is only weakly
dependent on its energy.  Above $1$\,TeV the resolution improves because
radiative energy losses become dominant.  Above $100$\,TeV the resolution
degrades, because energy loss fluctuations dominate.

Although these methods are quite different, their performances are
similar.  The full $E_{reco}$ and $P_{hit}$ methods achieve similar
resolutions up to $1\,$PeV.  The $P_{hit}$ method becomes worse above this
energy, because in \amax almost all of the OMs are hit, and the method
saturates.  In contrast, the Neural Net method shows a slightly poorer
resolution up to $1\,$PeV but is better above.  Its resolution is
relatively constant over several decades of energy.  This is an advantage
when reconstructing an original energy spectrum with an unfolding
procedure as in \cite{GEENEN//:2002A}.

\begin{figure}[htpb]
\centerline{
\epsfig{file=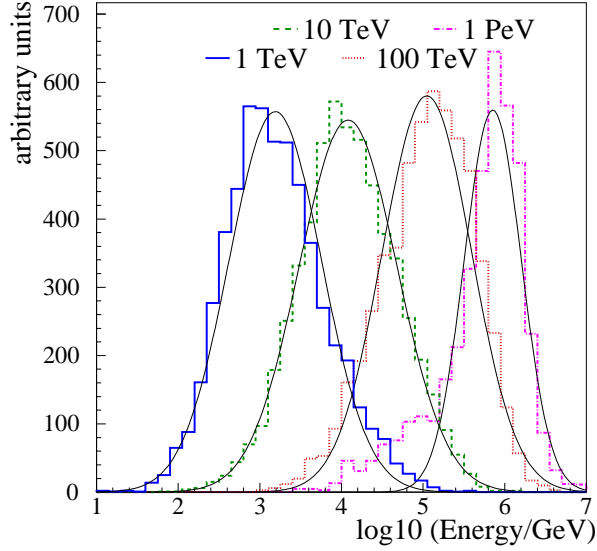,width=0.49\textwidth}
}
\caption{Energy reconstruction for simulated muons of different fixed energy  in \amaii,
using the neural net method. 
\label{fig:monoenergyres}}
\end{figure}

The \amaii detector contains more than twice as many OMs as \amax, and the
energy resolution is  better, especially at larger energies,
 $\sigma ( \Delta \log_{10} E ) \simeq 0.3 $.
The neural net reconstruction results for \amaii are shown in
figure~\ref{fig:monoenergyres}.
Finally, the recently installed transient waveform recorders (TWR) allow
 better amplitude measurements, which should significantly 
improve the results of
the energy reconstructions, in particular, the \emph{full $E_{reco}$}
method \cite{WAGNER//:2003:A}.

As discussed in section~\ref{sec:intro}, the cascade channel can achieve
substantially better resolutions, because the full energy is deposited inside
or close of the detector.
Energy resolutions in $\Delta \log_{10} E $ of $\le 0.2$ and $\le 0.15$ can be
achieved by \amax and \amaii, respectively \cite{AHRENS//:2002:D}.

\subsection{Systematic Uncertainties \label{sec:systematics}}

\begin{figure}[htpb]
\centerline{
\epsfig{file=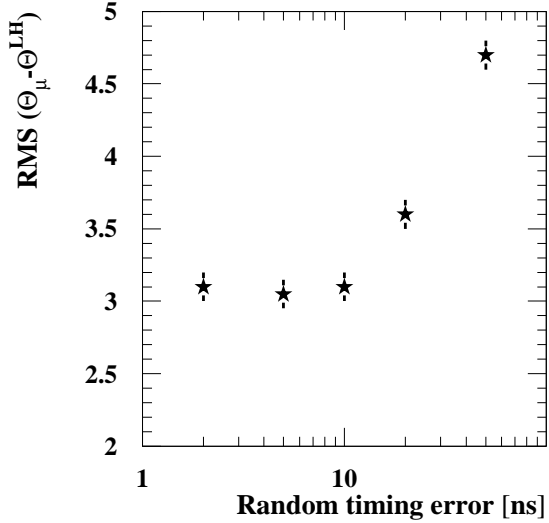,width=0.45\textwidth}}
\caption[Dependence of angular resolution on T0 uncertainty]
{The 
zenith angle deviations (RMS)  as function of an additional uncertainty
in the $t_0$ time calibration.
Data is shown for simulated atmospheric neutrino events in \amax
with the selection of \cite{AHRENS//:2002:C}.
The transit time of the PMTs has been shifted 
without correcting for in the reconstruction.
The shift is a fixed value for each PMT, obtained from a random Gaussian 
distribution.
\label{fig:res_timing}
}
\end{figure}

Several parameters of the detector are calibrated and therefore only known
with limited accuracy.  These parameters include the time offsets, the 
OM positions and the  absolute OM sensitivities.  We have
estimated the effects of these uncertainties on the resolution of \aman
reconstructions~\cite{BIRON:2002:A}.
As an example, figure~\ref{fig:res_timing} shows the effect of an
additional contribution to the time calibration uncertainty 
for the 10 string \amax detector.  The zenith angular resolutions for simulated
atmospheric neutrino events only degrade when the additional timing
uncertainties exceed $ 10\,$ns.  Additional tests with similar results were
done with non-random systematic shifts such as a depth dependent shift or a
string dependent shift.  Therefore, the angular resolution is insensitive
to the  uncertainties in the time calibration.  The geometry of the
detector is known to better than $30\,$cm horizontally and to better than
$1\,$m vertically, which corresponds to timing uncertainties of $\lesssim
1\mbox{ or } 3.5\,$ns, respectively.  Therefore, the geometry calibration
is also sufficiently accurate.

Similarly, the effect of uncertainties on other parameters, like the
absolute PMT efficiency, has been investigated.  No indication was found
that the remaining calibration uncertainties seriously 
affect the angular resolution
or the  systematic zenith angle offset.
  The combined calibration uncertainties are expected to
affect the accuracy of the reconstruction by less than $5\,$\% in the
zenith angle resolution and to less than $0.5^\circ$ in the absolute
pointing offset.

\section{Discussion and Outlook \label{sec:ideas}}

We have developed methods to reconstruct and identify muons induced by
neutrinos \cite{AHRENS//:2002:C}, inspite of the challenges of the natural
environment and large backgrounds.  These methods allow
us to establish \aman as a working neutrino telescope. 
The reconstruction techniques described in this paper are 
still subject to improvement in several aspects:
\begin{description}
\item[The likelihood description:] The likelihood functions for track
        reconstruction are based on the assumption of exactly one
        infinitely long muon track per event.  Extensions of this model to
        encompass \emph{starting muon tracks} (including the description of
        the hadronic vertex), \emph{stopping muons}, \emph{muon bundles} of
        non-negligible width, and \emph{multiple independent muons} will be
        important, particularly in the context of larger detectors such as
        Ice Cube.  Initial efforts fitting multiple muons with the direct
        walk algorithm have been  useful in rejecting coincident
        down-going muons, and work toward reconstructing muon bundles has
        begun in the context of events coincident with \spase air showers.

\item[The p.d.f.\ calculation:] The likelihood func\-tion is based on
        parametrizations of probability density functions (p.d.f.).  The
        p.d.f.\ is obtained from Monte Carlo simulations, and its accuracy is
        limited by the accuracy of the simulation.  Better simulations lead
        directly to a better p.d.f.\ and hence better reconstructions.

\item[The p.d.f.\ parametrizations:] The p.d.f.\ is parametrized by functions
        (e.g. the \emph{Pandel functions}) which only approximate the full
        p.d.f..  More accurate parametrization functions will result in
        better reconstructions.  For example, the scattering coefficient
        shows a significant depth dependence (see section~\ref{sec:aman}).
        The current reconstruction is based on an average p.d.f.\ assuming
        depth-independent ice properties.  While the track reconstruction
        is relatively insensitive to the accuracy of the parametrization,
        we expect a depth-dependent p.d.f.\ to have better energy
        reconstruction.

\item[Complementary information:] The current reconstruction algorithms do
        not include all available information in an event.  In
        particular, correlations between detected PMT signals are ignored.  For this
        reason dedicated selection parameters have been designed to
        exploit this information. They are used to discriminate between
        well reconstructed and poorly reconstructed events and improve the
        quality of the data sample.  Future work will try to improve these
        parameters and expand the present likelihood description.

\item[Transient waveform recorders:]
At the beginning of the year 2003, the detector readout has been
upgraded  with transient waveform recorders \cite{WAGNER//:2003:A}.
We expect a substantial improvement of the multiple-photon detection 
and  the dynamic range in particular for high muon energies.

\end{description}
 
The construction of a much larger detector, the \ice detector,
will start in the year 2004. It
will  consist of 4800 PMT deployed on 80 vertical strings and  
will surround the \aman detector \cite{YOSHIDA//:2003:A}.
The performance of \ice has been studied with 
realistic Monte Carlo simulations and 
similar analysis techniques as described in this paper \cite{AHRENS//:2003:B}.
The result is a substantially improved performance in terms
of sensitivity and reconstruction accuracy.
A direction accuracy of about $0.7^\circ $ (median) for energies above $1$\,TeV is
achieved. Similar to  \aman,  we expect a further
improvement by exploiting  the full information,
avaliable from the recorded wave-forms, in the reconstruction.

\paragraph*{Acknowledgments}
This research was supported by the following agencies: 
Deutsche Forschungsgemeinschaft (DFG);
German Ministry for Education and Research; 
Knut and Alice Wallenberg Foundation, Sweden; 
Swedish Research Council; 
Swedish Natural Science Research Council; 
Fund for Scientific Research (FNRS-FWO), Flanders
Institute to encourage scientific and technological research in
industry (IWT), and Belgian Federal Office for Scientific, Technical
and Cultural affairs (OSTC), Belgium.
%
UC-Irvine AENEAS Supercomputer Facility; 
University of Wisconsin Alumni Research Foundation; 
U.S. National Science Foundation, Office of Polar Programs; 
U.S. National Science Foundation, Physics Division; 
U.S. Department of Energy; 
D.F. Cowen acknowledges the support of the NSF CAREER program.
I. Taboada acknowledges the  support of FVPI.

\bibliographystyle{elsart-num}

\bibliography{ref,icrc26,icrc27}

\begin{thebibliography}{10}
\expandafter\ifx\csname url\endcsname\relax
  \def\url#1{\texttt{#1}}\fi
\expandafter\ifx\csname urlprefix\endcsname\relax\def\urlprefix{URL }\fi

\bibitem{ANDRES//:2001:A}
E.~Andr{\'e}s, et~al., {Observation of High Energy Neutrinos with \cer
  detectors embedded in deep Antarctic Ice}, Nature 410~(6827) (2001) 441--443.

\bibitem{LEARNED/MANNHEIM:2001:A}
J.~G. Learned, K.~Mannheim, {High-Energy Neutrino Astrophysics}, Annual Reviews
  of Nuclear and Particle Science 50 (2000) 679--749.

\bibitem{LEARNED/PAKVASA:1995:A}
J.~G. Learned, S.~Pakvasa, {Detecting Nutau Oscillations at PeV Energies},
  Astroparticle Physics 3 (1995) 267--274.

\bibitem{AHRENS//:2002:D}
J.~Ahrens, et~al., {Search for Neutrino-Induced Cascades with the \aman
  Detector}, Physical Review~D 67 (2003) 012003,
  \mbox{arXive:astro-ph}/0206487.

\bibitem{HUNDERTMARK//:2003:A}
{"Stephan Hundertmark}, {the \aman Collaboration}, {\amax Limit on UHE
  Muon-Neutrinos}, in: {Proceedings of the 28th International Cosmic Ray
  Conference}, Tsukuba, Japan, 2003.

\bibitem{HUNDERTMARK//:2001:A}
S.~Hundertmark, et~al., {A Method to Detect {UHE} Neutrinos with {\aman}}, in:
  {Proceedings of the $27^{th}$ International Cosmic Ray Conference}, Vol.~3,
  Hamburg, Germany, 2001, pp. 1129--1132, {HE.236}.

\bibitem{DICKINSON//:2000:A}
J.~E. Dickinson, et~al., {A new air-Cherenkov array at the South Pole}, Nuclear
  Instruments and Methods in Physics Research~A 440~(1) (2000) 95.

\bibitem{AHRENS//:2002:E}
J.~Ahrens, et~al., {Calibration and Survey of \aman with the \spase Detectors},
  accepted for publication in Astroparticle Physics.
\newline\urlprefix\url{http://www.amanda.uci.edu/documents.html}

\bibitem{BAI//:2001:A}
X.~Bai, et~al., {Calibration and Survey of {\aman} with {\spase}}, in:
  {Proceedings of the $27^{th}$ International Cosmic Ray Conference}, Vol.~3,
  Hamburg, Germany, 2001, pp. 977--980, {HE.208}.

\bibitem{RAWLINS:2001:A}
K.~Rawlins, {Measuring the Composition of Cosmic Rays with the {\spase} and
  \aman Detectors}, Ph.D. thesis, University of Wisconsin, Madison, Wi, USA
  (Oct. 2001).
\newline\urlprefix\url{http://amanda.berkeley.edu/manuscripts/}

\bibitem{ANDRES//:2000:A}
E.~Andr{\'e}s, et~al., {The \aman neutrino telescope: principle of operation
  and first results}, Astroparticle Physics 13~(1) (2000) 1--20,
  \mbox{arXive:astro-ph}/9906203.

\bibitem{COWEN//:2001:A}
D.~F. Cowen, K.~Hanson, {the {\aman} Collaboration}, {Time Calibration of the
  {\aman} Neutrino Telescope with Cosmic Ray Muons}, in: {Proceedings of the
  $27^{th}$ International Cosmic Ray Conference}, Vol.~3, Hamburg, Germany,
  2001, pp. 1133--1136, {HE.237}.

\bibitem{WOSCHNAGG//:1999:A}
K.~Woschnagg, et~al., {Optical Properties of South Pole Ice at Depths from 140
  to 2300 Meters}, in: D.~Kieda, M.~Salamon, B.~Dingus (Eds.), {Proceedings of
  the $26^{th}$ International Cosmic Ray Conference}, Vol.~2, Salt Lake City,
  USA, 1999, pp. 200--203, {HE.4.1.15}.

\bibitem{PRICE/WOSCHNAGG:2001:A}
P.~B. Price, K.~Woschnagg, {Role of Group and Phase Velocity in High-Energy
  Neutrino Observatories}, Astroparticle Physics 15~(1) (2001) 97--100,
  \mbox{arXive:hep-ex}/0008001.

\bibitem{PRICE/WOSCHNAGG:2001:B}
K.~Woschnagg, P.~B. Price, {Temperature Dependence of Absorption in Ice at
  532~nm}, Applied Optics 40~(15) (2001) 2496--2500.

\bibitem{PRICE/WOS/CHI:2000:A}
P.~B. Price, K.~Woschnagg, D.~Chirkin, {Age vs depth of glacial ice at South
  Pole}, Geophysical Research Letters 27~(13).

\bibitem{HE/PRICE/:1998:A}
Y.~D. He, P.~B. Price, {Remote Sensing of Dust in Deep Ice at the South Pole},
  Journal Geophysical Research~D 103~(14) (1998) 17041--17056.

\bibitem{PRICE/BERGSTOEM:1997:A}
P.~B. Price, L.~Bergst{\"o}m, {Optical Properties of Deep Ice at the South
  Pole: scattering}, Applied Optics 36~(18) (1997) 4181--4194.

\bibitem{ASKEBJER//:1997:A}
P.~Askebjer, et~al., {Optical Properties of Deep Ice at the South Pole:
  Absorption}, Applied Optics 36~(18) (1997) 4168--4180.

\bibitem{PDG:2000:A}
{Particle Data Group}, {Review of Particle Properties}, European Physical
  Journal~C {15}~(1--4).

\bibitem{KUZMICHEV:2000:A}
L.~A. Kuzmichev, {On the Velocity of Light Signals in the Deep Underwater
  Neutrino Experiments}, \mbox{e-preprint}, Moscow State University,
  \mbox{arXive:hep-ex}/0005036 (Mar. 2000).

\bibitem{SPIERING//:1993:A}
C.~Spiering, et~al., {Track Reconstruction and Background Rejection in the
  Baikal Neutrino Telescope}, in: {Proceedings of the 3rd Nestor International
  Workshop}, Pylos, Greece, 1993, p. 234, {DESY 94-050}.

\bibitem{CARMONA//:2001:A}
E.~Carmona, {Reconstruction Methods for the Antares Neutrino Telescope}, in:
  {Proceedings of the $2^{nd}$ Workshop on Methodical Aspects of Underwater/Ice
  Neutrino Telescopes}, Hamburg, Germany, 2001, p. 111.

\bibitem{BOUCHTA:1998:A}
A.~Bouchta, {Muon Analysis with the \amab four-string detector}, Ph.D. thesis,
  Stockholms Universitet, Stockholm, Sweden, uSIP Report 1998-07 (1998).
\newline\urlprefix\url{http://amanda.berkeley.edu/manuscripts/}

\bibitem{PORRATA:1997:A}
R.~A. Porrata, {The Energy Spectrum of Pointlike Events in \amaa}, Ph.D.
  thesis, University of California, Irvine, Ca, USA (1997).

\bibitem{MIOCINOVIC:2001:A}
P.~{Mio\v{c}inovi{\'c}}, {Muon energy reconstruction in the Antarctic Muon and
  Neutrino Detector Array (\aman)}, Ph.D. thesis, University of California,
  Berkeley, Ca, USA (Dec. 2001).
\newline\urlprefix\url{http://amanda.berkeley.edu/manuscripts/}

\bibitem{HILL:2001:A}
G.~C. Hill, {Bayesian event reconstruction and background rejection in neutrino
  detectors}, in: {Proceedings of the $27^{th}$ International Cosmic Ray
  Conference}, Vol.~3, Hamburg, Germany, 2001, pp. 1279--1282, {HE.267}.

\bibitem{Hill//:2002:A}
{Ty R. DeYoung and Gary C. Hill}, et~al., {Application of Bayes' Theorem to
  Muon Track Reconstruction in \aman}, in: {Proceedings Advanced Statistical
  Techniques in Particle Physics}, Durham, UK, 2002.

\bibitem{COUSINS//:2002:A}
R.~D. Cousins, {Conference summary talk}, in: {Proceedings Advanced Statistical
  Techniques in Particle Physics}, Durham, UK, 2002.

\bibitem{AHRENS//:2002:C}
J.~Ahrens, et~al., {Observation of High Energy Atmospheric Neutrinos with the
  Antarctic Muon and Neutrino Detector Array}, Physical Review~D 66~(1) (2002)
  012005, \mbox{arXive:astro-ph}/0205109.

\bibitem{DEYOUNG:2001:A}
T.~R. DeYoung, {Observation of Atmospheric Muon Neutrinos with \aman}, Ph.D.
  thesis, University of Wisconsin at Madison, Madison, Wisconsin, USA (May
  2001).
\newline\urlprefix\url{http://amanda.berkeley.edu/manuscripts/}

\bibitem{AYNUTDINOV//:2001:A}
V.~M. Aynutdinov, et~al., Physics of Particles and Nuclei Letters 109 (2001)
  43.

\bibitem{KARLE:1999:C}
A.~Karle, {Monte Carlo simulation of photon transport and detection in deep
  ice: muons and cascades}, in: {Proceedings of Workshop on the Simulation and
  Analysis Methods for Large Neutrino Telescopes}, DESY-Proc-1999-01, DESY
  Zeuthen, Germany, 1999, pp. 174--185.

\bibitem{WIEBUSCH:1999:A}
C.~H. Wiebusch, {Muon reconstruction with {\aman}}, in: {Proceedings of
  Workshop on the Simulation and Analysis Methods for Large Neutrino
  Telescopes}, DESY-Proc-1999-01, DESY Zeuthen, Germany, 1999, pp. 302--316.

\bibitem{PANDEL:1996:A}
D.~Pandel, {Bestimmung von Wasser- und Detektorparametern und Rekonstruktion
  von Myonen bis 100~TeV mit dem Baikal-Neutrinoteleskop NT-72}, Diploma
  thesis, Humboldt-Universit{\"a}t zu Berlin, Berlin, Germany (Feb. 1996).
\newline\urlprefix\url{http://www-zeuthen.desy.de/nuastro/publications/diploma%
/}

\bibitem{GEENEN//:2002A}
H.~Geenen, {Energy reconstruction and spectral unfolding of atmospheric leptons
  with the AMANDA detector}, Diploma thesis, University of Wuppertal,
  Wuppertal, Germany (Nov. {\\2002}).
\newline\urlprefix\url{{http://amanda.berkeley.edu/manuscripts}}

\bibitem{AHRENS//:2003:A}
J.~Ahrens, et~al., {Limits on diffuse fluxes of high energy extraterrestrial
  neutrinos with the {\amax} detector}, Physical Review Letters 90 (2003)
  251101, \mbox{arXive:astro-ph}/0303218.

\bibitem{STENGER:1990:A}
V.~J. Stenger, {Track fitting for {DUMAND}-{II} Octagon Array}, External Report
  HDC-1-90, University of Hawai'i at Manoa, Manoa, Hawaii, USA (1990).

\bibitem{PAW:1999:A}
{CERN Information Technology Division}, {``{PAW} -- Physics Analysis
  Workstation; User's Guide''}, CERN Program Library Long Writeup Q121, Geneva,
  Switzerland (1999).

\bibitem{JACOBSEN:1999:A}
J.~Jacobsen, C.~Wiebusch, {An Overview of Offline Software for {\aman}}, in:
  {Proceedings of Workshop on the Simulation and Analysis Methods for Large
  Neutrino Telescopes}, DESY-Proc-1999-01, DESY Zeuthen, Germany, 1999, pp.
  194--204.

\bibitem{STREICHER//:2001:C}
O.~Streicher, C.~Wiebusch, {``recoos''}, muon and neutrino reconstruction for
  underwater/ice Cherenkov telescopes (2001).
\newline\urlprefix\url{http://www.ifh.de/nuastro/software/siegmund/}

\bibitem{STREICHER//:2001:D}
O.~Streicher, C.~Wiebusch, {``rdmc''}, a library for processing neutrino
  telescope data and MC files (2001).
\newline\urlprefix\url{http://www.ifh.de/nuastro/software/siegmund/}

\bibitem{STREICHER//:2001:B}
G.~C. Hill, S.~Hundertmark, M.~Kowalski, P.~Miocinovic, T.~Neunh{\"o}fer,
  P.~Niessen, P.~Steffen, O.~Streicher, C.~Wiebusch, {``The Si{EGM}u{ND}
  software package''} (2001).
\newline\urlprefix\url{http://www.ifh.de/nuastro/software/siegmund/}

\bibitem{PRESS/TE/VE/FL:1997:A}
W.~H. Press, S.~A. Teukolsky, W.~V. Vetterling, B.~P. Flannery, {``Numerical
  Recipies in {C} -- The Art of Scientific Computing''}, 2nd Edition, Cambridge
  University Press, Cambridge, UK, 1997.
\newline\urlprefix\url{http://www.nr.com/}

\bibitem{JAMES:1994:A}
F.~James, {the CERN Computing and Networks Division}, {``{MINUIT} -- Function
  Minimization and Error Analysis; Reference Manual''}, CERN Program Library
  Long Writeup D506, Geneva, Switzerland, 94th Edition (1994).

\bibitem{YOUNG:2001:A}
S.~Young, {A Search for Point Sources of High Energy Neutrinos with the \amax
  Neutrino Telescope}, Ph.D. thesis, University of California at Irvine,
  Irvine, California, USA (Jul. 2001).
\newline\urlprefix\url{http://amanda.berkeley.edu/manuscripts/}

\bibitem{AHRENS//:2002:F}
J.~Ahrens, et~al., {Search for Point Sources of High Energy Neutrinos with
  \aman}, Astrophysical Journal 583 (2003) 1040,
  \mbox{arXive:astro-ph}/0208006.

\bibitem{BOESER:2002:A}
S.~B{\"o}ser, {Separation of atmospheric neutrinos with the \amaii detector},
  Diploma thesis, Technische Universit{\"a}t M{\"u}nchen, Munich, Germany (Apr.
  2002).

\bibitem{AHRENS//:2002:B}
J.~Ahrens, et~al., {Limits to the muon flux from WIMP annihilation in the
  center of the Earth with the \aman detector}, Physical Review~D 66~(3) (2002)
  032006, \mbox{arXive:astro-ph}/0202370.

\bibitem{NIESSEN//:2001:A}
P.~Niessen, C.~Spiering, et~al., {Search for Relativistic Monopoles with the
  {\aman} Detector}, in: {Proceedings of the $27^{th}$ International Cosmic Ray
  Conference}, Vol.~4, Hamburg, Germany, 2001, pp. 1496--1498, {HE.315}.

\bibitem{BIRON:1998:A}
A.~Biron, {On the Rejection of Atmospheric Muons in the {\aman} Detector},
  Diploma thesis, Humboldt-Universit{\"a}t zu Berlin, Berlin, Germany,
  \mbox{DESY-THESIS-1998-014} (Mar. {\\1998}).
\newline\urlprefix\url{http://amanda.berkeley.edu/manuscripts/}

\bibitem{GAUG:2001:A}
M.~Gaug, {AMANDA event reconstruction and cut evaluation methods}, in:
  R.~Wischnewski (Ed.), {Proceedings of the $2^{nd}$ Workshop on Methodical
  Aspects of Underwater/Ice Neutrino Telescopes}, Hamburg, Germany, 2001, p.
  123.

\bibitem{GAUG:2000:A}
M.~Gaug, {Detection of Atmospheric Muon Neutrinos with the {\aman} Neutrino
  Telescope}, Diploma thesis, Humboldt-Universit{\"a}t zu Berlin, Berlin,
  Germany (Oct. 2000).
\newline\urlprefix\url{http://amanda.berkeley.edu/manuscripts/}

\bibitem{GAUG:2000:B}
M.~Gaug, {\cuteval}, Website, including manual (2000).
\newline\urlprefix\url{{http://www-zeuthen.desy.de/$\sim$gaug/cuteval/}}

\bibitem{LEUTHOLD:2001:A}
M.~J. Leuthold, {Search for Cosmic High Energy Neutrinos with the {\amax}
  Detector}, Ph.D. thesis, Humboldt-Universit{\"a}t zu Berlin, Berlin, Germany
  (Sep. 2001).

\bibitem{BIRON:2002:A}
A.~Biron, {Search for Atmospheric Muon-Neutrinos and Extraterrestric Neutrino
  Point Sources in the 1997 \amax Data}, Ph.D. thesis, Humboldt-Universit{\"a}t
  zu Berlin, Berlin, Germany (Jan. {2002}).
\newline\urlprefix\url{http://amanda.berkeley.edu/manuscripts/}

\bibitem{WAGNER//:2003:A}
{Wolfgang Wagner}, {the \aman Collaboration}, {New Capabilities of the
  AMANDA-II High Energy Neutrino Telescope}, in: {Proceedings of the 28th
  International Cosmic Ray Conference}, Tsukuba, Japan, 2003.

\bibitem{YOSHIDA//:2003:A}
{Shigeru Yoshida}, {the \ice Collaboration}, {The IceCube High Energy Neutrino
  Telescope}, in: {Proceedings of the 28th International Cosmic Ray
  Conference}, Tsukuba, Japan, 2003.

\bibitem{AHRENS//:2003:B}
J.~Ahrens, et~al., {Sensitivity of the IceCube Detector to Astrophysical
  Sources of High Energy Muon Neutrinos}, submitted for publication to
  Astroparticle Physics, \mbox{arXive:astro-ph}/0305196 (May 2003).

\end{thebibliography}

\end{document}